\documentclass[ 12pt, a4paper]{article}

\usepackage[utf8x]{inputenc}
\usepackage{amssymb,amsmath,graphicx,epsfig,tikz,setspace,stmaryrd,mathtools}

\makeatletter
\newsavebox\myboxA
\newsavebox\myboxB
\newlength\mylenA

\newcommand*\xoverline[2][0.75]{%
    \sbox{\myboxA}{$\m@th#2$}%
    \setbox\myboxB\null
    \ht\myboxB=\ht\myboxA%
    \dp\myboxB=\dp\myboxA%
    \wd\myboxB=#1\wd\myboxA
    \sbox\myboxB{$\m@th\overline{\copy\myboxB}$}
    \setlength\mylenA{\the\wd\myboxA}
    \addtolength\mylenA{-\the\wd\myboxB}%
    \ifdim\wd\myboxB<\wd\myboxA%
       \rlap{\hskip 0.5\mylenA\usebox\myboxB}{\usebox\myboxA}%
    \else
        \hskip -0.5\mylenA\rlap{\usebox\myboxA}{\hskip 0.5\mylenA\usebox\myboxB}%
    \fi}
\makeatother

\renewcommand{\vec}[1]{\boldsymbol{#1}}
\newcommand{\be}{\begin{equation}}
\newcommand{\en}{\end{equation}}
\newcommand{\bb}{\mathcal B}

\def\bga#1\ega{\begin{gather}#1\end{gather}} 
\def\bgas#1\egas{\begin{gather*}#1\end{gather*}}

\def\bal#1\eal{\begin{align}#1\end{align}} 
\def\bals#1\eals{\begin{align*}#1\end{align*}}

\def \ini#1{\overset{\mathbf{I}}{ #1}}

\DeclareMathOperator{\tr}{tr}

\begin{document}
\title{\sc A new restriction for initially stressed elastic solids}
\author{
	Artur L. Gower$^1$, Tom Shearer$^1$ and Pasquale Ciarletta$^2$ \\[12pt]
	\footnotesize{$^1$ School of Mathematics, University of Manchester, Oxford Road, Manchester M13 9PL, UK}\\
	\footnotesize{$^2$ MOX- Politecnico di Milano, piazze Leonardo da Vinci 32, 20133 Milano, Italy}
}
\date{\today}
\maketitle

\begin{abstract}
We introduce a fundamental restriction on the strain energy function and stress tensor for initially stressed elastic solids. The restriction applies to strain energy functions $W$ that are explicit functions of the elastic deformation gradient $\mathbf{F}$ and initial stress $\vec \tau$, i.e. $W:= W(\mathbf F, \vec \tau)$. The restriction is a consequence of energy conservation and ensures that the predicted stress and strain energy do not depend upon an arbitrary choice of reference configuration. We call this restriction \emph{initial stress reference independence} (ISRI). It transpires that \textit{almost all} strain energy functions found in the literature do not satisfy ISRI, and may therefore lead to unphysical behaviour, which we illustrate via a simple example. To remedy this shortcoming we derive three strain energy functions that \emph{do} satisfy the restriction. We also show that using \emph{initial strain} (often from a virtual configuration) to model initial stress leads to strain energy functions that automatically satisfy ISRI. Finally, we reach the following important result: ISRI reduces the number of unknowns of the linear stress tensor of initially stressed solids. This new way of reducing the linear stress may open new pathways for the non-destructive determination of initial stresses via ultrasonic experiments, among others.
\end{abstract}





\section{Introduction}
Materials in many contexts operate under a significant level of internal stress, often called \emph{residual stress} if the material is not subjected to any external loading. Residual stress is desirable in many circumstances; for example, living matter uses residual stress to preserve ideal mechanical conditions for its physiological function \cite{fung_what_1991,holzapfel_biomechanics_2003}. In manufacturing, if residual stress is controlled, it can be used to strengthen materials such as turbine blades \cite{james2010shot} and toughened glass \cite{todd1999thermal}; however, residual stress is often problematic as it can cause materials to fail prematurely~\cite{webster_residual_2001,lennon_residual_2002}. \emph{Pre-stress} is another common term, which refers to internal stress caused by an external load~\cite{parnell2012employing,parnell2013antiplane,shearer2013torsional,shearer2015antiplane}. In this paper, the term initial stress is used to describe any internal stress, irrespective of boundary conditions, and therefore encompasses \emph{both} residual stress \emph{and} pre-stress.

In both industrial and biological contexts, the origin and extent of initial stresses are often unknown. One way to determine these stresses is by measuring how they affect the elastic response of the material. In metallurgy, it is well known that residual stress can be estimated by drilling small holes into a metal and observing how they change shape~\cite{rossini_methods_2012}. Elastic waves are also used in many applications, since their behaviour is very sensitive to the initial stress in a material \cite{guz2002elastic}.

One alternative to link the response of the material to a very general dependence on the internal stress, therefore including initially stressed materials, is the implicit form of elasticity described by Rajagopal and coworkers~\cite{bustamante2011solutions,rajagopal_implicit_2003,rajagopal_response_2007}, but this generality comes with the drawback of adding greater constitutive complexity.  Explicit hyperelastic models are simpler and are accurate for many applications -- the work of Hoger~\cite{hoger_determination_1986,hoger_elasticity_1993} and  Man~\cite{man_towards_1987,man_hartigs_1998} has led to improved inverse methods for measuring initial stress~\cite{robertson_determining_1998,rachele_uniqueness_2003,lin_uniqueness_2003,alessandrini_inverse_2003,sharafutdinov_tomography_2012,joshi_reconstruction_2013} and monitoring techniques~\cite{castellano_monitoring_2016}.

The mechanical properties of a hyperelastic material can be conveniently determined from its strain energy function $W$, which gives the strain energy per unit volume of the initially stressed \textit{reference} configuration. In classical elasticity, $W$ is a function of only the elastic deformation gradient $\mathbf{F}$ (i.e. $W:=W(\mathbf{F})$). The simplest way to account for initial stresses is to allow $W$ to depend on either the initial Cauchy stress tensor $\vec \tau$, or on an initial deformation gradient $\mathbf{F}_0$ from some stress-free configuration $\mathcal{B}_0$. For the first method, $W := W(\mathbf{F},\vec\tau)$~\cite{guillou_growth_2006,shams_wave_2010,shams_initial_2011}, whereas for the second, $W := J_0^{-1}W_0(\mathbf{F}\mathbf{F}_0)$~\cite{hoger_elasticity_1993,johnson_use_1995}, where $J_0=\det\mathbf{F}_0$ and $W_0$ is the strain energy per unit volume in $\mathcal{B}_0$. In both cases, $\mathbf F$ is the elastic deformation gradient from the initially stressed to the current configuration.

The two approaches each have relative advantages and disadvantages. If measuring the initial stress is the main goal, then using $W:= W(\mathbf{F},\vec \tau)$ is the more direct method, but requires an extra restriction -- ISRI (presented below). It is also the more useful form when the initial stress is postulated or known a priori, such as assuming that the stress gradient in an arterial wall tends to be homogeneous ~\cite{gower_initial_2015}. If $W := J_0^{-1}W_0(\mathbf{F}\mathbf{F}_0)$, then the classical theory of nonlinear elasticity can be used (by taking $\mathcal{B}_0$ as the reference configuration), and ISRI is automatically satisfied. This form is more useful when a stress-free configuration is known, or when the exact form of the initial stress is not important. The two approaches are not equivalent
because it is not always possible to deduce  $\mathbf F_{0}$ from $\vec \tau$, as they are related by the equilibrium equation of the initially stressed configuration, which is a nonlinear partial differential equation in $\mathbf F_0$. We discuss initially strained models in Section~\ref{sec:InitialStrain}.

The primary purpose of this paper is to deduce a fundamental restriction on $W:=W(\mathbf F, \vec \tau)$, and discuss its consequences. To motivate the need for a new restriction, we show how a simple uniaxial deformation can lead to unphyscial results when this restriction is ignored in Section~\ref{sec:uniaxial}. In Section~\ref{sec:restrict}, we derive this restriction, which follows from the fact elastic deformations conserve energy and we call it \emph{initial stress reference independence} (ISRI), for reasons that will be clarified later. We assume the only source of anisotropy is due to the initial stress, though a more general form of ISRI could also be deduced for materials that include other sources of anistropy. ISRI can be stated solely in terms of stress tensors, and should therefore hold for materials whose constitutive behaviour is not expressed in terms of a strain energy function.

It transpires that it is not easy to choose a strain energy function that satisfies ISRI. In fact, almost every strain energy function used in the literature to date does not satisfy it, in both finite elasticity~\cite{shams_initial_2011,merodio_influence_2013,shams_rayleigh-type_2014,merodio_extension_2015,shams_effect_2016,nam_effect_2016} and linear elasticity~\cite{man_hartigs_1998,shams_wave_2010}. To the authors' knowledge, the only existing strain energy function that does satisfy ISRI is that derived in \cite{gower_initial_2015}, which is an initially stressed incompressible neo-Hookean solid, as discussed in Section~\ref{DoesSatisfy}. To address this lack of valid models, we present two new strain energy functions that satisfy ISRI in Section~\ref{sec:newmodels}. In Section~\ref{sec:InitialStrain}, we discuss strain energy functions based on \emph{initial strain}, and show that they automatically satisfy ISRI in Section~\ref{sec:InitialStrainImpliesISO}.

Small elastic deformations on initially stressed solids lead to easier connections between the elastic response and the initial stress. This makes them ideal for establishing methods to measure initial stress. An important consequence of ISRI is that it restricts the linearised elastic stress tensor $\delta \vec \sigma(\mathbf F, \vec \tau)$, as we discuss in Section \ref{sec:IncrementalISO}. For materials subjected to \textit{small} initial stress, we use ISRI to reduce the number of unknowns in $\delta \vec \sigma(\mathbf F, \vec \tau)$ in Section~\ref{sec:ModerateInitialStress}. The result is a reduced version of the stress tensor deduced in~\cite{man_hartigs_1998}, which could ultimately improve the measurement of initial stress via ultrasonic experiments, among others.

In the literature, it is common to deduce the linear stress tensor $\delta \vec \sigma$ by considering an initial strain from a stress free configuration~\cite{thurston_third-order_1964, gandhi_acoustoelastic_2012, kubrusly_derivation_2016}. This approach is broadly called acousto-elasticity, and as discussed in Section~\ref{sec:InitialStrain}, the resulting $\delta \vec \sigma$ automatically satisfies ISRI, but leads to an indirect connection between $\delta \vec \sigma$ and $\vec \tau$. In fact, acousto-elasticity was used by Tanuma and Man \cite{tanuma_perturbation_2008} to restrict the form of $\delta \vec \sigma(\mathbf F, \vec \tau)$ when both strain and initial stress are small, which led them to our equation~\eqref{eqns:Alpha8Restriction} (their equation (81)). In our approach we clarify that this equation must hold for every initially stressed elastic material, regardless of the origins of this stress.

\section{Initial stress reference independence}
\label{sec:ISO}

The mechanical properties of an elastic material can be determined from its strain energy function $W$, which gives the strain energy per unit volume of the reference configuration. For an initially stressed material, $W$ can be expressed in terms of the deformation gradient $\mathbf{F}$ from the reference to the current configuration and $\vec \tau$, the Cauchy stress in the reference configuration, so that $W:= W (\mathbf{F}, \vec \tau)$. In general, $W$ may also depend on position, but we omit this dependency for clarity. We call $\vec \tau$ the initial stress tensor and, when discussing consititutive choices, we will not require any specific boundary conditions in the reference configuration, in agreement with \cite{merodio_influence_2013} (i.e the boundaries can either be loaded or unloaded).

In what follows, we assume that $\mathbf{F}$ is within the \textit{elastic} regime of the material, but make no assumptions about how the initial stress formed. The Cauchy stress tensor ${\vec \sigma}$ \cite{ogden_non-linear_1997,guillou_growth_2006} for an initially stressed material is given by
\be
\vec{\sigma} := \vec{\sigma}(\mathbf F,\vec \tau) =  J^{-1}\mathbf{F}\frac{\partial W}{\partial\mathbf{F}}(\mathbf{F},\vec \tau)-p\mathbf{I},
\label{eqn:CauchyStress}
\en
where $J=\det\mathbf{F}$, $\mathbf{I}$ is the identity tensor and $p$ is zero if the material is compressible or, otherwise, is a Lagrange multiplier associated with the incompressibility  constraint $\det \mathbf{F}=1$. We define differentiation with respect to a second-order tensor as follows:
\begin{equation}
\left(\frac{\partial}{\partial\mathbf{A}}\right)_{ij}=\frac{\partial}{\partial\text{A}_{ji}}.
\end{equation}

Before moving on, we present an example where a specific choice of $W(\mathbf F, \vec \tau)$ leads to two different stress responses for the same uniaxial deformation.

\subsection{Motivating example}
\label{sec:uniaxial}

\begin{figure}[ht!]
\centering{
\includegraphics[width=0.85\textwidth]{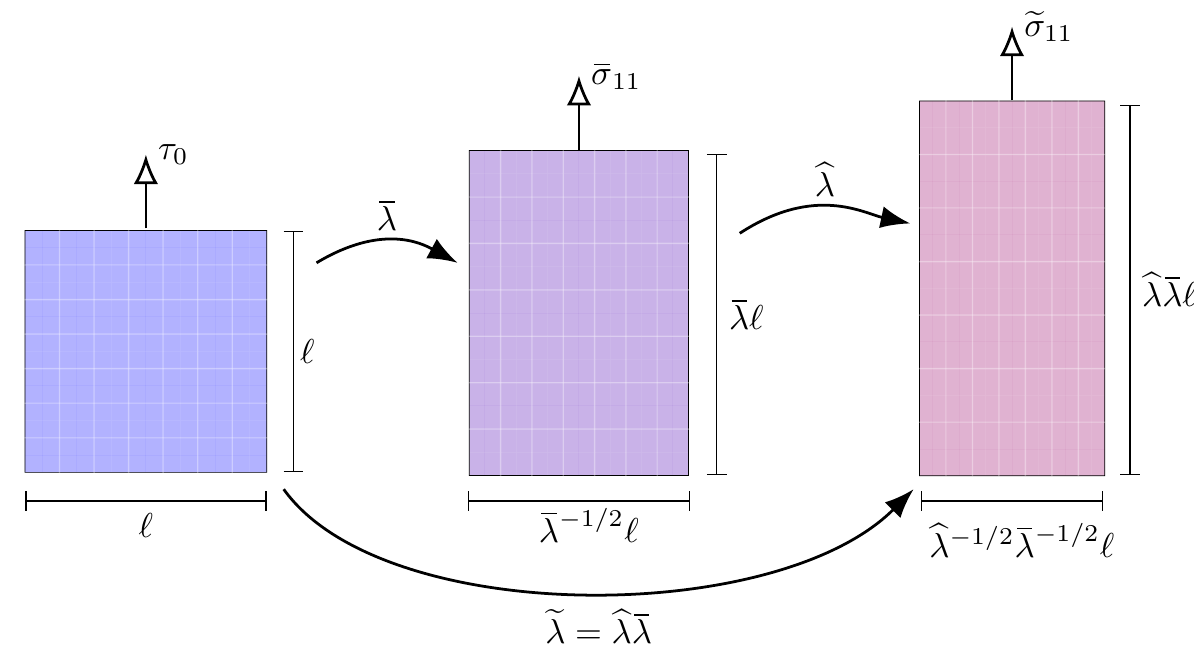}
\vspace{-.2cm}
}
\caption{Uniaxial deformation of an initially stressed cube (depth not illustrated), with sides of length $\ell$, into a cuboid of height $\widehat \lambda \xoverline \lambda \ell$ and width (equal to depth) $\widehat \lambda^{-1/2} \xoverline \lambda^{-1/2} \ell$. The hollow arrows represent the stress applied to the top boundary. The uniaxial stretch $\widetilde \lambda$ is indicated by the bottom arrow. This stretch can also be achieved in two steps: first a stretch of $\xoverline \lambda$, then, a further stretch of $\widehat \lambda$. The second of these stretches treats the middle configuration as its reference configuration. Both of these ways of achieving the same uniaxial stretch $\widehat \lambda \xoverline \lambda$ should require the same stress $\widetilde \sigma_{11}$ in the rightmost configuration.}
\label{fig:uniaxial}
\end{figure}

To study the influence of initial stress on the elastic response of a material, a simple strain energy function was postulated by Merodio \textit{et al.} \cite{merodio_influence_2013} as follows
\begin{equation}
W_\text{MOR}=\frac{\nu}{2}\left(\tr (\mathbf F^{\text{T}} \mathbf F)-3 \right)+\frac{1}{2} \left ( \tr (\mathbf F^{\text{T}} \vec \tau \mathbf F)  - \tr \vec \tau \right),
\label{eqn:W_MOR}
\end{equation}
where $\nu$ is a mateiral constant, the superscript $\text{T}$ indicates the transpose operator and $\tr$ the trace. As $W_\text{MOR}$ is used for incompressible materials, the Cauchy stress~\eqref{eqn:CauchyStress} becomes
\be
\vec \sigma = - p \mathbf I +\nu \mathbf F \mathbf F^\text{T}  + \mathbf F \vec \tau \mathbf F^\text{T}.
\label{eqn:merodioStress}
\en
Consider an initially stressed material described by Euclidean coordinates $(X,Y,Z)$. Suppose the initial stress takes the form of a homogeneous tension $T$ along the $X$ axis, and that the material is subsequently stretched along the same axis, then the components of the deformation gradient and initial stress tensor are given by
\be{}
	\text{F} =
	\begin{pmatrix}
	\lambda & 0 & 0 \\
	0 & \lambda^{-1/2} & 0 \\
	0 & 0 & \lambda^{-1/2} \\
	\end{pmatrix}
	\quad \text{and} \quad{}
	\tau =
	\begin{pmatrix}
	T & 0 & 0 \\
	0 & 0 & 0 \\
	0 & 0 & 0 \\
	\end{pmatrix},
	\label{eqns:uniaxial}
\en
where $\lambda$ is the amount of stretch. Applying stress-free boundary conditions on the faces not under tension gives $p = \lambda^{-1} \nu$, which in turn leads to
\be
 \sigma_{11} := \sigma_{11}(\lambda, T) = \lambda^{2} (\nu + T)  -\lambda^{-1} \nu,
\label{eqn:ex_BC}
\en
which is the stress necessary to support any stretch $\lambda$ given an initial tension $T$. We will now choose two different ways of achieving the same uniaxial stretch $\lambda = \widetilde \lambda$ that should, \textit{but do not}, result in the same stress when using the strain energy function \eqref{eqn:W_MOR}. First, we consider a direct application of the stretch $\lambda = \widetilde \lambda$ and assume that the initial tension is $T =\tau_{0}$. In this case,
\be
\widetilde \sigma_{11} = \sigma_{11}(\widetilde \lambda, \tau_{0}) = \widetilde \lambda^{2} (\nu + \tau_{0})  - \widetilde \lambda^{-1} \nu.
\label{eqn:example_sigma11}
\en
We can also achieve the same stretch in two steps by taking $\widetilde \lambda = \widehat \lambda \xoverline \lambda$. That is, first we stretch by $\xoverline \lambda$ and then apply a further stretch $\widehat \lambda$, as shown in Figure~\ref{fig:uniaxial}. Taking $\lambda = \xoverline \lambda$, and again using $T = \tau_{0}$, results in the stress
\be
\xoverline \sigma_{11} = \sigma_{11}( \xoverline \lambda, \tau_{0}) =  \xoverline \lambda^{2} (\nu + \tau_{0})  - \xoverline \lambda^{-1} \nu,
\en
in the intermediate configuration. To further stretch the material, we take this intermediate configuration as our intially stressed reference configuration, where the initial tension is now $T =\xoverline \sigma_{11}$. Upon applying the second stretch $\widehat \lambda$, we obtain
\bal
	\widetilde \sigma_{11} &=  \sigma_{11}(\widehat \lambda, \xoverline \sigma_{11}) = \widehat \lambda^{2}  (\nu + \xoverline \sigma_{11})  - \widehat \lambda^{-1} \nu
	\\
	&=  \widehat \lambda^{2} \xoverline \lambda^{2} (\nu + \tau_{0}) +\widehat \lambda^{2} \nu  - \widehat \lambda^{2} \xoverline \lambda^{-1} \nu  - \widehat \lambda^{-1} \nu.
	\label{eqn:example2_sigma11}
\eal
Both \eqref{eqn:example_sigma11} and \eqref{eqn:example2_sigma11} result from the same uniaxial deformation, so should be identical, but, upon substituting $\widetilde \lambda = \widehat \lambda \xoverline \lambda$ into \eqref{eqn:example_sigma11}, we find they are not.

If, instead of equation~\eqref{eqn:merodioStress}, we had used an initially strained model, e.g. an incompressible neo-Hookean $W := \mu \tr (\mathbf F \mathbf F_0)/2$, then this unphysical result would not occur. However, as explained in the introduction, when the initial strain/stress are unknown, both $\vec \tau$ and $\mathbf F_0$ are unknown, and an explicit form $W: = W(\mathbf F, \vec \tau)$ leads to more direct connections between the elastic response and initial stress $\vec \tau$.

The unphysical behaviour illustrated by this example is typical of many of the strain energy functions of the form $W := W(\mathbf F, \vec \tau)$ in the literature and highlights the need to restrict what forms of $W(\mathbf{F},\vec\tau)$ are physically permissible. Therefore, in the following section, we present a restriction on $W(\mathbf F, \vec \tau)$ that ensures that such unphysical behaviour does not occur.

\subsection{The restriction}
\label{sec:restrict}

The elastic energy stored in a material should remain constant under a rigid motion, so $W(\mathbf F, \vec \tau) = W (\mathbf Q \mathbf F, \vec \tau)$ for every proper orthogonal tensor $\mathbf Q$ (so that
$\mathbf Q \mathbf Q^\text{T} =\mathbf I$ and $\det\mathbf{Q}=1$). This identity can be used to show that $W$ depends on $\mathbf F$ only through the right Cauchy-Green tensor $\mathbf C = \mathbf F^\text{T} \mathbf F$~\cite{ogden_non-linear_1997}, which we use to rewrite the Cauchy stress~\eqref{eqn:CauchyStress} as
\be
\vec{\sigma}(\mathbf F,\vec \tau) = 2J^{-1}\mathbf{F}\frac{\partial W}{\partial\mathbf{C}}(\mathbf{C},{\vec \tau})\mathbf{F}^\text{T}-p\mathbf{I}.
\label{eqn:CauchyStressC}
\en

\begin{figure}[ht!]
\centering{
\begin{tikzpicture} [scale=1.]
 	\draw (3.2,-3.7) node {\includegraphics[width=0.32\textwidth]{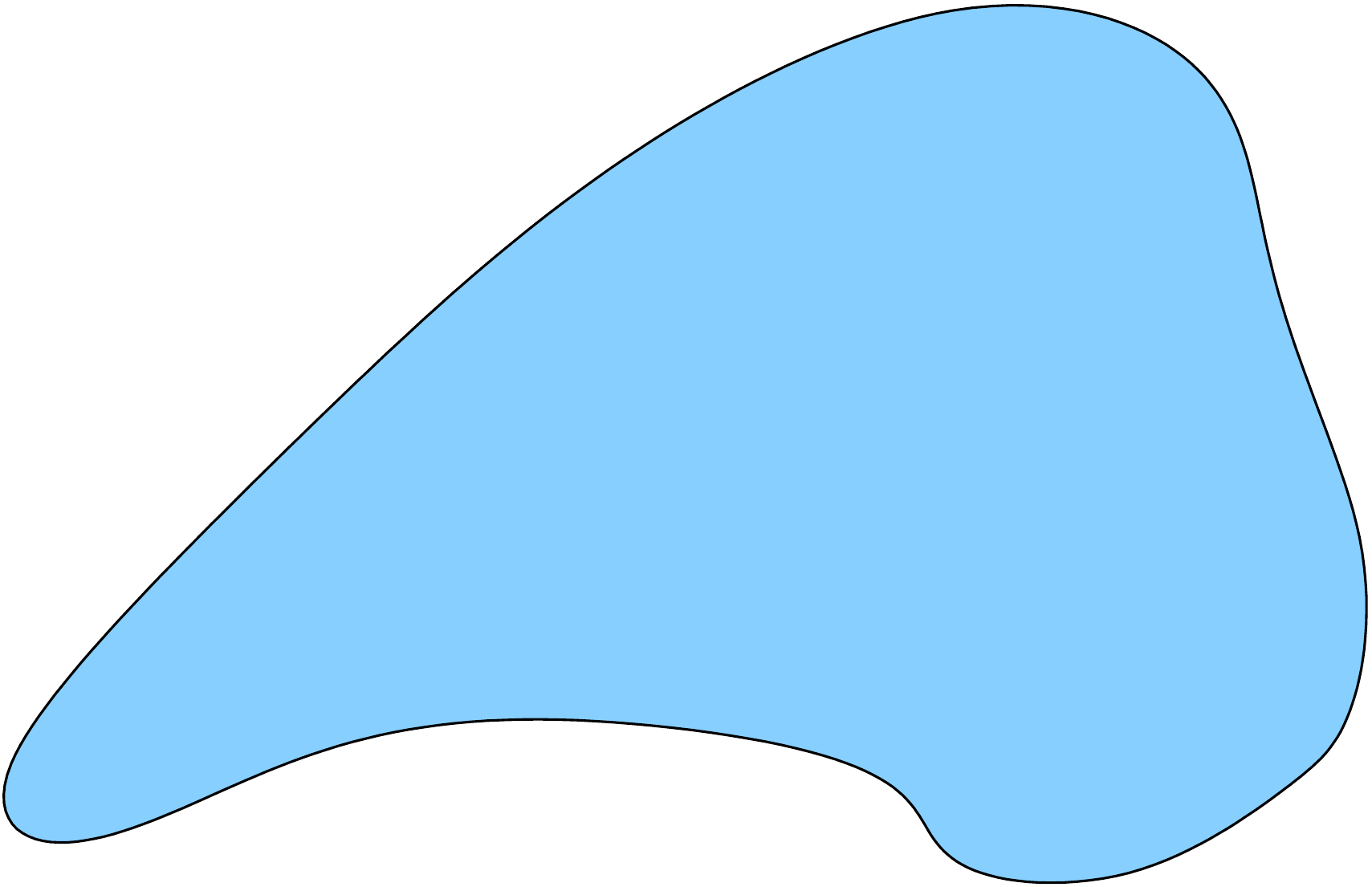}};
 	\draw (2.8,-2.5) node {$\xoverline{\bb}$};
 	\draw (-1,0) node {\includegraphics[width=0.3\textwidth]{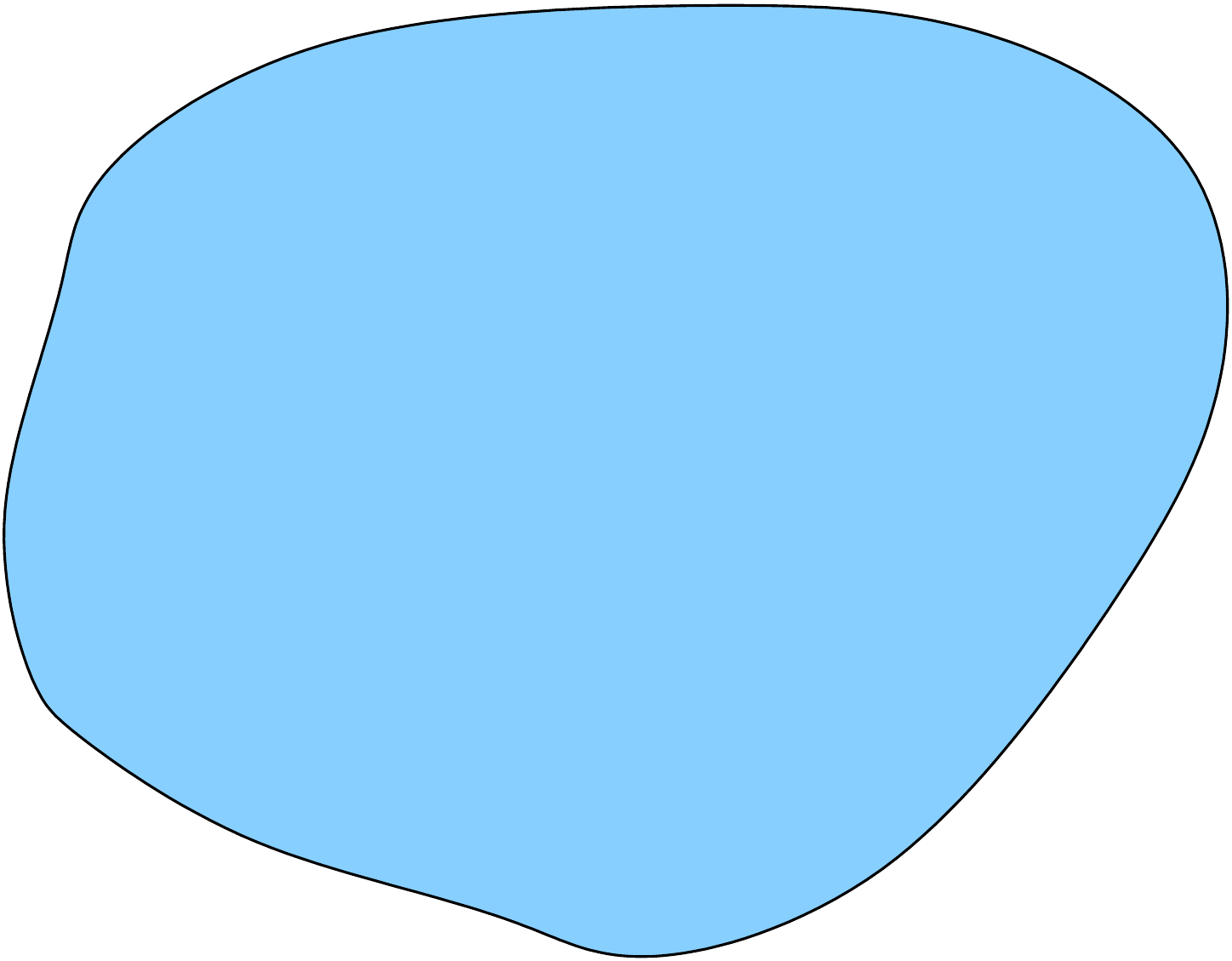}};
 	\draw (-2.6,1.5) node {$\bb$} ;
 	\draw (7,0) node {\includegraphics[width=0.25\textwidth]{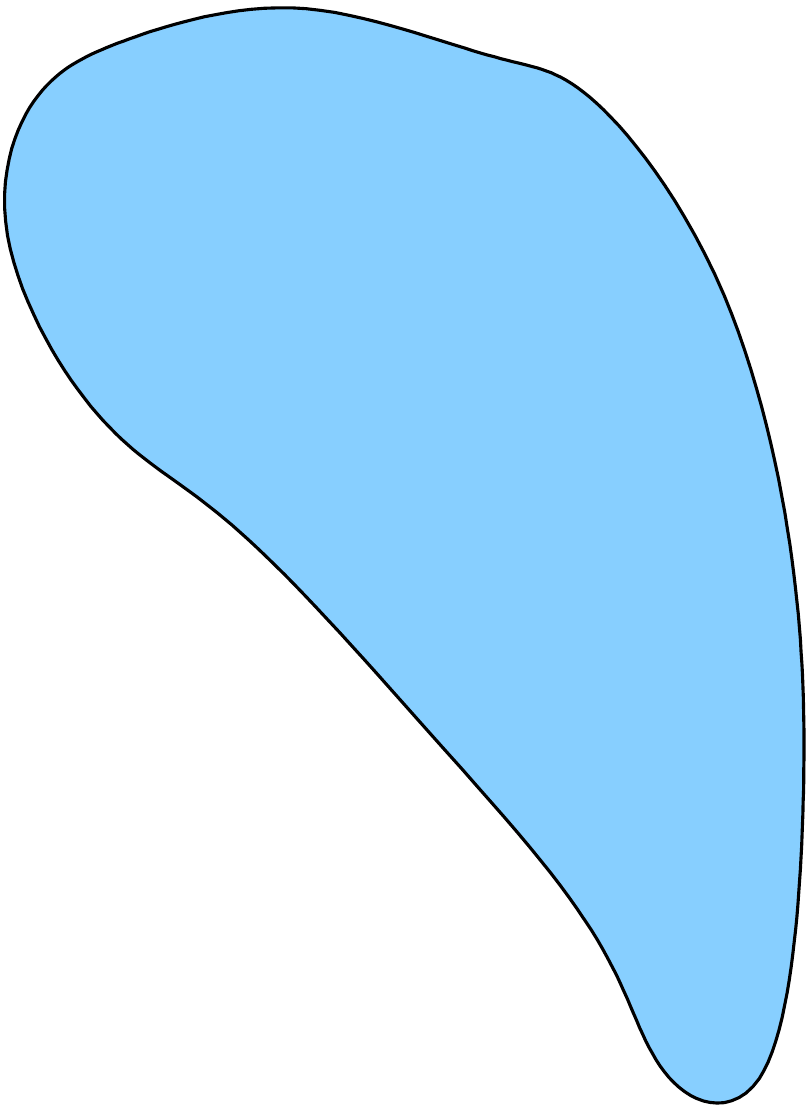}};
 	\draw (8.5,1.6) node {$\widetilde{\bb}$} ;
 	\draw (7.3,0.7) node {$\widetilde{\boldsymbol{\sigma}}$};
 	\draw (-1,0.1) node{$\boldsymbol{\tau}$};
 	\draw (3.9,-3.6) node{$\xoverline{\boldsymbol{\sigma}}$};
 	\draw [->] (1.,1.2) .. controls (2.,1.6) and (3,1.9) .. (5.2,1.3)[thick];
 	\draw (3,1.7) node[above]{$\widetilde{\mathbf{F}}=\widehat{\mathbf{F}}\xoverline{\mathbf{F}}$};
 	\draw [->] (0.,-1.4) .. controls (0.3,-2.5) and (1,-3.) .. (1.95,-3.55)[thick];
 	\draw (0.35,-2.5) node[below]{$\xoverline{\mathbf{F}}$};
 	\draw [->] (5.4,-3.65) .. controls (6.,-3.5) and (7.,-3.) .. (7.5,-1.5)[thick];
 	\draw (6.6,-3.2) node[right]{$\widehat{\mathbf{F}}$};
\end{tikzpicture}
}
\caption{Deformation of initially stressed elastic solids. The stress and strain energy in $\widetilde{\bb}$ should not depend on whether $\bb$ or $\xoverline{\bb}$ is taken as the reference configuration.}
\label{fig:Configurations}
\end{figure}

The presence of initial stress generally leads to an anisotropic material response, but for simplicity we assume that no other source of anisotropy is present. Referring to the three configurations shown in Figure~\ref{fig:Configurations}, let the strain energy per unit volume in $\widetilde \bb$ be denoted by $\psi$. The strain energy due to the elastic deformation from $\bb$ to $\widetilde \bb$ should be the same as that due to successive elastic deformations from $\bb$ to $\xoverline \bb$, then from $\xoverline \bb$ to $\widetilde \bb$. In detail, taking $\bb$ as the reference configuration, we conclude $\psi = \widetilde {J}^{-1}  W( \widehat{\mathbf F} \xoverline{\mathbf F}, \vec \tau)$ where $\widetilde J= \widehat J \, \xoverline J$, $\widehat{J}=\det\widehat{\mathbf{F}}$ and $\xoverline{J}=\det\xoverline{\mathbf{F}}$, whereas if $\xoverline \bb$ is taken as the reference configuration, we conclude $\psi =  \widehat{J}^{-1} W(\widehat{\mathbf F}, \vec \sigma(\xoverline{\mathbf F}, \vec \tau ) )$. Since these two quantities must be equal, we therefore have
\be
\boxed{
W(\widehat{\mathbf F} \xoverline{\mathbf F}, \vec \tau) = \xoverline J W \left ( \widehat{\mathbf F}, {\vec \sigma}(\xoverline {\mathbf F}, \vec \tau ) \right ) \;\; \text{for every} \;\; \vec \tau, \, \xoverline {\mathbf F} \; \text{and} \; \widehat {\mathbf F}
}
\label{eqn:ISO}
\en
where both $\xoverline {\mathbf F}$ and $\widehat{\mathbf F}$ are associated with \textit{elastic} deformations (which may be constrained by incompressibility). We refer to this criterion as \emph{initial stress reference independence} (ISRI). The vast majority of initially stressed strain energy functions in the literature~\cite{shams_wave_2010,shams_initial_2011,merodio_influence_2013,shams_rayleigh-type_2014,merodio_extension_2015,shams_effect_2016,nam_effect_2016} do not satisfy this restriction and therefore may exhibit physically unrealistic behaviour.

When $\xoverline{\mathbf F}= \mathbf I$, equation~\eqref{eqn:ISO} reduces to $W (\widehat {\mathbf F}, \vec \tau) =W (\widehat {\mathbf F}, {\vec \sigma}(\vec I, \vec \tau))$, which, from equation~\eqref{eqn:CauchyStressC}, is always satisfied if
\be
{\vec \sigma}(\vec I, \vec \tau)=2\frac{\partial W}{\partial\mathbf{C}}(\mathbf{I},{\vec \tau})-p\mathbf{I}=\vec{\tau},
\label{eqn:StressCompatibility}
\en
for every $\vec \tau$. We refer to this well known restriction as \textit{initial stress compatibility}. Additionally, if $\mathbf F= \mathbf Q$, where again $\mathbf{Q}$ is a proper orthogonal tensor representing a rigid body motion, then, using equations~\eqref{eqn:CauchyStressC} and~\eqref{eqn:StressCompatibility}, we reach ${\vec \sigma}(\mathbf Q, \vec \tau) = \mathbf Q \vec \tau \mathbf Q^\text{T}$. Using this result, along with $\xoverline{\mathbf{F}}=\mathbf{Q}$ in equation~\eqref{eqn:ISO}, we obtain
\be
	W(\widetilde {\mathbf F}, \vec \tau) = W \left ( \widetilde {\mathbf F} \mathbf Q^\text{T},\mathbf Q \vec \tau \mathbf Q^\text{T}  \right ),
	\label{eqn:RigidObjectivity}
\en
where $\widetilde {\mathbf F} = \widehat{ \mathbf F} \xoverline {\mathbf F}$. The above identity is typically used for anisotropic materials~\cite{spencer_theory_1971} and can be used to derive the following ten independent invariants~\cite{shams_initial_2011}\footnote{Note that the invariants $I_{\vec{\tau}_1}$, $I_{\vec{\tau}_2}$ and $I_{\vec{\tau}_3}$ are different from, but can expressed as linearly independent combinations of, those derived in~\cite{shams_initial_2011}.}
\bal
	& I_1= \tr {\mathbf C}, \;\;\;\; I_2=\frac{1}{2}[(I_1^2 -\tr({\mathbf C }^2)], \;\;\;\;
	I_3= \det {\mathbf C}
	\label{eqns:Is},
	\\
	& I_{\vec{\tau}_1}= \tr \vec \tau, \;\;\; I_{\vec{\tau}_2}=\frac{1}{2}[(I_{\vec{\tau}_1}^2 -\tr({\vec \tau }^2)], \;\;
	I_{\vec{\tau}_3}= \det (\vec \tau)
	\label{eqns:Itau},
	\\
	& J_1= \tr (\vec \tau{\mathbf C}), \;\; J_2= \tr (\vec \tau {\mathbf C}^2), \;\;  J_3= \tr ({\vec
	\tau}^2 {\mathbf C}), \;\; J_4= \tr (\vec \tau^2 {\mathbf C}^2).
	\label{eqns:Icombined}
\eal
Using these invariants, the Cauchy stress can be rewritten as
\begin{multline}
\boldsymbol{\sigma}(\mathbf{F},\boldsymbol{\tau})=-p \mathbf I +\frac{1}{J}\left(2W_{I_1}\mathbf{B}+2W_{I_2}(I_1\mathbf{B}-\mathbf{B}^2)+\right.\\
\left.2I_3W_{I_3}\mathbf{I}+2W_{J_1}\mathbf{F}\vec{\tau}\mathbf{F}^\text{T}+2W_{J_2}(\mathbf{F}\vec{\tau}\mathbf{F}^\text{T}\mathbf{B}+\mathbf{BF}\vec{\tau}\mathbf{F}^\text{T})+2W_{J_3}\mathbf{F}\boldsymbol{\tau}^2\mathbf{F}^\text{T}+\right.\\
\left.2W_{J_4}(\mathbf{F}\boldsymbol{\tau}^2\mathbf{F}^\text{T}\mathbf{B}+\mathbf{BF}\boldsymbol{\tau}^2\mathbf{F}^\text{T})\right),
\label{sigmacomp}
\end{multline}
where $\mathbf{B}=\mathbf{FF}^\text{T}$ is the left Cauchy-Green tensor, $W_{I_i}=\partial W/\partial I_i$ and $W_{J_j}=\partial W/\partial J_j$, with $i\in\{1,2,3\}$ and $j\in\{1,2,3,4\}$. For an incompressible material $I_3=1$ and $W_{I_3} =0$. Note that the Cauchy stress in a standard non-linear elastic material can be obtained from \eqref{sigmacomp} simply by letting $W$ depend only on the \textit{strain} invariants $I_1$, $I_2$ and $I_3.$

By evaluating equation~\eqref{sigmacomp} at $\mathbf F=\mathbf I$ we obtain another form of the initial stress compatibility equation~\eqref{eqn:StressCompatibility}:
\begin{multline}
\vec \tau =  \mathbf I ( - \ini p +2 \ini W_{I_1} +4 \ini W_{I_2}+2 \ini W_{I_3})\\
+\vec \tau (2 \ini W_{J_1} +4 \ini W_{J_2}) +\vec \tau^2 (2 \ini W_{J_3} +4 \ini W_{J_4}),
\end{multline}
where the notation $\ini{\cdot}$ is used to denote that $\cdot$ is evaluated at $\mathbf F= \mathbf I$ \textit{after} differentiation. Since this equation has to hold for \textit{any} initial stress tensor $\vec \tau$, the initial stress compatibility condition is equivalent to
\be
2 \ini W_{I_1} +4 \ini W_{I_2} +2 \ini W_{I_3} =\ini p, \quad   2\ini W_{J_1} +4\ini W_{J_2}=1, \quad \ini W_{J_3} +2 \ini W_{J_4} =0.
\label{eqn:StressCompatibilityScalar}
\en
In the literature, $W$ is often chosen as a simple function of the ten invariants~(\ref{eqns:Is},\ref{eqns:Icombined}) that satisfy initial stress compatibility~\eqref{eqn:StressCompatibilityScalar}. However, it is highly unlikely that any $W$ chosen {\it a priori} will satisfy ISRI~\eqref{eqn:ISO}.

A version of ISRI can also be stated in terms of the stress tensor, without reference to a strain energy function. To do so, assume the internal stress is given by some constitutive choice $\vec \sigma := \vec \sigma(\mathbf F, \vec \tau)$. Then using reasoning similar to that which led to equation~\eqref{eqn:ISO} we find that
\begin{equation}
\boxed{
\boldsymbol{\sigma}(\widehat{\mathbf{F}}\xoverline{\mathbf{F}},\boldsymbol{\tau})=\boldsymbol{\sigma}(\widehat{\mathbf{F}},\boldsymbol{\sigma}(\xoverline{\mathbf{F}},\boldsymbol{\tau})), \;\; \text{for every} \; \vec \tau, \; \xoverline {\mathbf F} \; \text{and} \;  \widehat {\mathbf F}.}
\label{eqn:ISO2}
\end{equation}
This restriction states that the Cauchy stress in $\widetilde{\mathcal{B}}$ should not change when a different reference configuration is selected. By choosing $\widehat {\mathbf F} \xoverline {\mathbf F}= \mathbf I$ and using equation~\eqref{eqn:StressCompatibility}, we obtain $\vec \tau = {\vec \sigma}(\xoverline {\mathbf F}^{-1},\xoverline {\vec \sigma})$, where $\xoverline{\vec \sigma} =\vec \sigma(\xoverline{\mathbf F},\vec \tau)$. This restriction was derived in~\cite{gower_initial_2015} and termed \textit{initial stress symmetry}. It allowed a straightforward way to model the adaptive remodelling of living tissues such as arterial walls towards an ideal target stress~\cite{chuong_residual_1986,taber_stress-modulated_2001}. For more details see~\cite{gower_initial_2015} and \cite{ciarletta_morphology_2016}.

As demonstrated in Section~\ref{sec:uniaxial}, strain energy functions that do not satisfy ISRI may exhibit unphysical behaviour. To the authors' knowledge, the only strain energy function in the literature to date that does satisfy ISRI is that derived in \cite{gower_initial_2015}. We prove this in the following section, then derive two new strain energy functions that satisfy ISRI in Section~\ref{sec:newmodels}.

\subsection{An incompressible strain energy function that satisfies ISRI}
\label{DoesSatisfy}

In a recent paper, Gower \textit{et al.} \cite{gower_initial_2015} proposed the strain energy function
\begin{equation}
 W_\text{GCD}=\frac{1}{2}(p_0(I_{\boldsymbol{\tau}_1},I_{\boldsymbol{\tau}_2},I_{\boldsymbol{\tau}_3})I_1+J_1-3\mu),
\label{WGCD}
\end{equation}
where $p_0$ is a function of $I_{\boldsymbol{\tau}_1}$, $I_{\boldsymbol{\tau}_2}$ and $I_{\boldsymbol{\tau}_3}$ given by
\begin{equation}
p_0=\frac{1}{3}\left(T_3+\frac{T_1}{T_3}-I_{\boldsymbol{\tau}_1}\right),
\label{eqn:p0}
\end{equation}
\bal
	T_1 &=I_{\vec{\tau}_1}^2-3I_{\vec{\tau}_2}, \quad T_2=I_{\vec{\tau}_1}^3-\frac{9}{2}I_{\vec{\tau}_1}I_{\vec{\tau}_2}+\frac{27}{2}(I_{\vec{\tau}_3}-\mu^3),
	\\
	 T_3 &=\sqrt[3]{\sqrt{T_2^2-T_1^3}-T_2}.
\eal
One way to derive $W_\text{GCD}$ is to rewrite an initially strained neo-Hookean strain energy function as an initially stressed strain energy function~\cite{gower_initial_2015}. An alternative derivation is given in Appendix~\ref{app:DeducingValidW}. Using $W_\text{GCD}$ in equation \eqref{sigmacomp}, the left side of equation \eqref{eqn:ISO2} becomes
\begin{equation}
	\boldsymbol{\sigma}(\widehat{\mathbf{F}}\xoverline{\mathbf{F}},\boldsymbol{\tau})=p_0\widehat{\mathbf{F}}\xoverline{\mathbf{B}}\widehat{\mathbf{F}}{}^\text{T}-\widetilde{p}\,\mathbf{I}+\widehat{\mathbf{F}}\xoverline{\mathbf{F}}\boldsymbol{\tau}\xoverline{\mathbf{F}}{}^\text{T}\widehat{\mathbf{F}}{}^\text{T},
	\label{GCDstress1}
\end{equation}
and the right side becomes
\begin{equation}
\boldsymbol{\sigma}(\widehat{\mathbf{F}},\boldsymbol{\sigma}(\xoverline{\mathbf{F}},\boldsymbol{\tau}))=(p_1-\xoverline{p})\widehat{\mathbf{B}}+p_0\widehat{\mathbf{F}}\xoverline{\mathbf{B}}\widehat{\mathbf{F}}{}^\text{T}-\widehat{p}\,\mathbf{I}+\widehat{\mathbf{F}}\xoverline{\mathbf{F}}\boldsymbol{\tau}\xoverline{\mathbf{F}}{}^\text{T}\widehat{\mathbf{F}}{}^\text{T},
\label{GCDstress2}
\end{equation}
where $p_1$ is the Lagrange multiplier associated with $\xoverline{\mathbf F}$. In the Appendix~\ref{app:DeducingValidW} we show that $\xoverline{p}=p_1$, and therefore equation~\eqref{GCDstress2} reduces to
\begin{equation}
\boldsymbol{\sigma}(\widehat{\mathbf{F}},\boldsymbol{\sigma}(\xoverline{\mathbf{F}},\hat{\boldsymbol{\tau}}))=p_0\widehat{\mathbf{F}}\xoverline{\mathbf{B}}\widehat{\mathbf{F}}^\text{T}-\widehat{p}\,\mathbf{I}+\widehat{\mathbf{F}}\xoverline{\mathbf{F}}\boldsymbol{\tau}\xoverline{\mathbf{F}}^\text{T}\widehat{\mathbf{F}}^\text{T}.
\label{GCDstress3}
\end{equation}
ISRI~\eqref{eqn:ISO2} then states that
\begin{equation}
\boldsymbol{\sigma}(\widehat{\mathbf{F}}\xoverline{\mathbf{F}},\boldsymbol{\tau})=\boldsymbol{\sigma}(\widehat{\mathbf{F}},\boldsymbol{\sigma}(\xoverline{\mathbf{F}},\boldsymbol{\tau}))~~~~~\Leftrightarrow~~~~~\widehat{p}=\widetilde{p}.
\end{equation}
Since equations \eqref{GCDstress1} and \eqref{GCDstress3} have exactly the same functional form and they must be subjected to the same boundary conditions because they both represent the Cauchy stress in $\widetilde{\bb}$, their Lagrange multipliers must be equal (i.e.\ $\widehat{p}=\widetilde{p}~$). Therefore, $W_\text{GCD}$ \textit{does} satisfy ISRI.

\subsection{Two compressible strain energy functions that satisfy ISRI}
\label{sec:newmodels}
By using the same method as that used in Appendix~\ref{app:DeducingValidW} to derive $W_\text{GCD}$, we have derived two new strain energy functions for compressible materials. Both are based on compressible extensions of the neo-Hookean model:
\begin{equation}
 W_\text{CNH1}=\frac{\mu}{2}(I_1-3-2\log\sqrt{I_3})+\frac{\lambda}{2}(\log\sqrt{I_3})^2,
\end{equation}
and
\begin{equation}
 W_\text{CNH2}=\frac{\mu}{2}(I_1-3-2\log\sqrt{I_3})+\frac{\lambda}{2}(\sqrt{I_3}-1)^2,
\end{equation}
where $\mu$ and $\lambda$ are the ground state first and second Lam\'e parameters, respectively. The initially stressed strain energy functions corresponding to these are
\begin{equation}
 W_\text{GSC1}=\frac{q_1}{2} I_1+\frac{J_1}{2}-\frac{\mu}{2 K_1}\left(3 + 2 \log(K_1 \sqrt{I_3}) \right) +\frac{\lambda}{2 K_1}\left(\log(K_1 \sqrt{I_3}) \right)^2
\end{equation}
and
\begin{equation}
 W_\text{GSC2}=\frac{q_2}{2}I_1+\frac{J_2}{2}-\frac{\mu}{2K_2}\left(3+2\log(K_2\sqrt{I_3})\right)+\frac{\lambda}{2K_2}\left(K_2\sqrt{I_3}-1\right)^2,
\end{equation}
where $q_1$, $q_2$, $K_1$ and $K_2$ are functions of $I_{\boldsymbol{\tau}_1}$, $I_{\boldsymbol{\tau}_2}$ and $I_{\boldsymbol{\tau}_3}$ and can be thought of as \textit{initial stress parameters} defined implicitly by the equations
\begin{equation}
 \frac{\mu^3}{K_1}=q_1^3+q_1^2I_{\boldsymbol{\tau}_1}+q_1I_{\boldsymbol{\tau}_2}+I_{\boldsymbol{\tau}_3},\qquad q_1=\frac{1}{K_1}(\mu-\lambda\log K_1),
\label{eqn:K1}
\end{equation}
\begin{equation}
 \frac{\mu^3}{K_2}=q_2^3+q_2^2I_{\boldsymbol{\tau}_1}+q_2I_{\boldsymbol{\tau}_2}+I_{\boldsymbol{\tau}_3},\qquad q_2=\frac{\mu}{K_2}+\lambda(1- K_2),
\label{eqn:K2}
\end{equation}
where the solutions for $K_1$ and $K_2$ should both be real and such that $K_1 \to 1$ and $K_2 \to 1$ when $\vec \tau \to \mathbf{0}$. The Cauchy stress tensors corresponding to these strain energy functions are, respectively,
\begin{equation}
\vec{\sigma}_\text{GSC1}=\frac{1}{J}\left(q_1\mathbf{B}+\frac{1}{K_1}(\lambda\log(JK_1)-\mu)\mathbf{I}+\mathbf{F}\vec{\tau}\mathbf{F}^\text{T}\right),
\label{incNH1}
\end{equation}
and
\begin{equation}
\vec{\sigma}_\text{GSC2}=\frac{1}{J}\left(q_2\mathbf{B}+\left(\lambda(I_3K_2-J)-\frac{\mu}{K_2}\right)\mathbf{I}+\mathbf{F}\vec{\tau}\mathbf{F}^\text{T}\right).
\label{incNH2}
\end{equation}
These constitutive equations provide a simple way to study the effects of initial stress on any deformation.

\section{Initially strained materials}
\label{sec:InitialStrain}
Another way to model initial stress is via initial strain. This is normally done by including an initial deformation gradient $\mathbf F_0$ from some configuration $\mathcal{B}_0$ in the strain energy function $W:= J_0^{-1} W_0(\mathbf F \mathbf F_0)$, where $J_0 =\det \mathbf F_0$ and $W_0$ is the strain energy per unit volume in $\mathcal{B}_0$. This representation of $W$ is a consequence of both a fundamental covariance argument~\cite{marsden_mathematical_1994,lu_covariant_2012}, and utilising a virtual stress-free configuration~\cite{johnson_use_1995}. The Cauchy stress tensor is then given by~\cite{marsden_mathematical_1994,lu_covariant_2012}
\be
\vec{\sigma} := \vec{\sigma}(\mathbf F \mathbf F_0) =  J^{-1}J_0^{-1}\mathbf{F}\frac{\partial W_0}{\partial\mathbf{F}}(\mathbf F \mathbf F_0)-p\mathbf{I}.
\label{eqn:CauchyStress2}
\en
Usually, $W_0(\mathbf F \mathbf F_0)$ is chosen such that $\mathcal{B}_0$ is stress-free, that is, $\vec \sigma (\mathbf I)= \vec 0$. Assuming that the initial strain is the only source of anisotropy, the strain energy can be shown to depend only on the isotropic invariants of $\mathbf F_0^\text{T}\mathbf C \mathbf F_0$:
\begin{equation}
 \widehat{I}_1=\tr(\mathbf{F}_0^\text{T}\mathbf{CF}_0),\quad\widehat{I}_2=\frac{1}{2}(\widehat{I}_1^2-\tr((\mathbf{F}_0^\text{T}\mathbf{CF}_0)^2)),\quad\widehat{I}_3=\det(\mathbf{F}_0^\text{T}\mathbf{CF}_0),
\end{equation}
so that $W:= J_0^{-1}W_0(\widehat{I}_1,\widehat{I}_2,\widehat{I}_3)$. These strain energy functions automatically satisfy ISRI, as shown below in Section~\ref{sec:InitialStrainImpliesISO}. An example of such a strain energy function is this initially strained form of the Mooney-Rivlin strain energy function:
\be
W_0 = C_1(\widehat{I}_1\widehat{I}_3^{\ -1/3}-3)+C_2(\widehat{I}_2\widehat{I}_3^{\ -2/3}-3) + C_3(\widehat{I}_3^{\ -1/2}-1)^2,
\en
where $C_1$, $C_2$ and $C_3$ are material constants that must be chosen such that the body is stress free when $\mathbf F= \mathbf F_0 =\mathbf I$.

Taking $W$ as a function of $\mathbf F$ and $\vec \tau$, or of $\mathbf F$ and $\mathbf F_0$, gives two different perspectives on the same phenomenon, each being useful in different circumstances. The former is more useful when the initial \textit{stress} is known, whereas the latter is more useful when the initial \textit{strain} can somehow be inferred.

\subsection{All initially strained materials satisfy ISRI}
\label{sec:InitialStrainImpliesISO}

We have discussed, in previous sections, that it is not easy to choose a function of the form $W:= W(\mathbf{F},\vec \tau)$ that satisfies ISRI~\eqref{eqn:ISO}. Let us consider the case of initially strained materials with
\be
W= W(\mathbf F, \vec \tau):= J_0^{-1} W_0(\mathbf F \mathbf F_0), \quad \text{and} \quad \vec \tau = \vec \sigma(\mathbf F_0).
\en
We will prove that if $W =W(\mathbf F, \vec \tau)$ is defined as above, it satisfies ISRI for \textit{any} choice of $W_0(\mathbf F \mathbf F_0)$. First we assume that for any $W_0$ and initial stress $\vec \tau$ there is a deformation gradient $\mathbf F_0$\footnote{Note that for there to be a unique $\mathbf F_0$, for every $\vec \tau$, some restrictions need to be made about the reference configuration of $\mathbf F_0$, see \cite{johnson_use_1995}, for example.} such that
\be
\vec \tau = \vec \sigma(\mathbf F_0) \quad \text{where} \quad \vec \sigma(\mathbf F_0) = J_0^{-1}\mathbf F_0  \frac{\partial W_0(\mathbf F_0)}{\partial \mathbf F_0}-p\mathbf{I}.
\en
Next, we define an initially \textit{stressed} strain energy function
\be
W(\mathbf{F},\vec\tau)=W(\mathbf{F},\vec \sigma(\mathbf F_0)):=J_0^{-1} W_0(\mathbf{FF}_0) \quad \text{for every} \;\; \mathbf F \;\; \text{and} \;\; \vec \tau.
\label{eqn:initstressinitstrain}
\en
By substituting $\mathbf{F}=\widehat{\mathbf{F}}\xoverline{\mathbf{F}}$ into equation \eqref{eqn:initstressinitstrain} we obtain
\begin{multline}
 W(\widehat{\mathbf{F}}\xoverline{\mathbf{F}},\vec\tau)= J_0^{-1} W_0((\widehat{\mathbf{F}}\xoverline{\mathbf{F}})\mathbf{F}_0)\\
= \xoverline J \xoverline J^{-1} J_0^{-1} W_0(\widehat{\mathbf{F}}(\xoverline{\mathbf{F}}\mathbf{F}_0))=\xoverline J W(\widehat{\mathbf{F}},\vec \sigma(\xoverline{\mathbf{F}}\mathbf F_0)).
\label{eqn:VirtualImpliesObjective}
\end{multline}
Then, using equation \eqref{eqn:CauchyStress2}, we obtain
\begin{equation}
\vec{\sigma}(\xoverline{\mathbf F} \mathbf F_0) =  \xoverline J^{-1} J_0^{-1}\xoverline{\mathbf{F}}\frac{\partial W_0}{\partial\xoverline{\mathbf{F}}}(\xoverline{\mathbf F} \mathbf F_0)-p\mathbf{I},
\end{equation}
and, since $J_0^{-1}W_0(\xoverline{\mathbf{F}}\mathbf{F}_0)=W(\xoverline{\mathbf{F}},\vec\tau)$,
\begin{equation}
\vec{\sigma}(\xoverline{\mathbf F} \mathbf F_0) =  \xoverline J^{-1}\xoverline{\mathbf{F}}\frac{\partial W}{\partial\xoverline{\mathbf{F}}}(\xoverline{\mathbf{F}},\vec\tau)-p\mathbf{I},
\end{equation}
which, using equation \eqref{eqn:CauchyStress}, gives
\begin{equation}
\vec{\sigma}(\xoverline{\mathbf F} \mathbf F_0) =\vec\sigma(\xoverline{\mathbf{F}},\vec\tau).
\end{equation}
Substituting the above into equation \eqref{eqn:VirtualImpliesObjective} we obtain $W(\widehat{\mathbf F} \xoverline{\mathbf F}, {\vec \tau}) = \xoverline J W(\widehat{\mathbf F},\vec \sigma( \xoverline{\mathbf F}, {\vec \tau}))$, which is the ISRI restriction~\eqref{eqn:ISO}.

Whilst such strain energy functions are guaranteed to satisfy ISRI, it is not often possible to state their dependence on the stress invariants $I_{\boldsymbol{\tau}_1}$, $I_{\boldsymbol{\tau}_2}$ and $I_{\boldsymbol{\tau}_3}$ \textit{explicitly} (a notable exception being the strain energy function discussed in Section~\ref{DoesSatisfy}). Instead, it may be necessary to define that dependence \textit{implicitly}, as is the case for the two models presented in Section~\ref{sec:newmodels}.

\section{Linear elasticity with initial stress}
\label{sec:IncrementalISO}

Elastic waves in solids are highly sensitive to initial stress, and linear elastic models fit measurements from currently employed experimental techniques well. Our aim here is, in the long run, to improve these measurements by using a linearised version of ISRI~\eqref{eqn:ISO}.

In Section~\ref{sec:IncrementalStress} we deduce the linearised stress without considering ISRI. Then, in Section~\ref{sec:GeneralIncremental}, we calculate a linearised form of ISRI and discuss how to use it to restrict the linearised stress. Hoger \cite{hoger_determination_1986,hoger_residual_1993}, Man and coworkers~\cite{man_towards_1987,man_hartigs_1998} derived the equations for small initial stress, up to first order in $\vec \tau$. In ~\cite{man_hartigs_1998} the authors remark that many experiments indicate that for small deformations the elastic stress depends linearly on the initial stress, at least for metals. Motivated by these observations, we linearise the elastic stress in both the elastic strain and initial stress in Section~\ref{sec:ModerateInitialStress} and reach a reduced form for the stress \eqref{eqn:LineariseStressAndTau} which adds a restriction to all previous models, to the authors' knowledge. The restriction \eqref{eqns:Alpha8Restriction} has been used before in the literature (see equation (81) from \cite{tanuma_perturbation_2008}) but was deduced from the context of acousto-elasticity.

\subsection{Linear elastic stress}
\label{sec:IncrementalStress}
For a small elastic deformation, we can write the associated deformation gradient as $\mathbf F = \mathbf I + \nabla\mathbf u$, where $\mathbf u$ is a small displacement. By Taylor series expanding the Cauchy stress \eqref{eqn:CauchyStress} about $\mathbf{F}=\mathbf{I}$, the linearised Cauchy stress becomes
\be
\delta\vec \sigma(\mathbf F, \vec \tau) = \vec \tau + \ini{\frac{\partial \vec \sigma}{\partial \mathbf F}}  :\nabla \mathbf u + \mathcal O((\nabla\mathbf u)^2),
\label{eqn:LineariseStress}
\en
where we have exploited the fact that $\vec\sigma(\mathbf{I},\vec{\tau})=\vec{\tau}$ and we remind the reader that $\ini{\cdot}$ denotes that $\cdot$ is evaluated at $\mathbf F= \mathbf I$ \textit{after} differentiation. We define
\begin{equation}
\left(\frac{\partial\mathbf{A}}{\partial\mathbf{B}}\right)_{ijkl}=\frac{\partial {A}_{ij}}{\partial {B}_{lk}}\qquad\text{and}\qquad(\vec{\mathcal{C}}:\mathbf{A})_{ij}=\mathcal{C}_{ij\alpha\beta} {A}_{\beta\alpha},
\label{eqn:Contraction}
\end{equation}
for any second-order tensors $\mathbf{A}$ and $\mathbf{B}$ and fourth-order tensor $\vec {\mathcal{C}}$, using Einstein summation convention for the repeated dummy indices $\alpha$ and $\beta$. Using equations~\eqref{eqn:CauchyStressC} and~\eqref{eqn:Contraction} it can be shown that
\begin{multline}
\ini{\frac{\partial \vec \sigma}{\partial \mathbf F}}: \mathbf A
= \frac{\partial}{\partial \mathbf F}\left.\left(2 J^{-1} \mathbf F \frac{\partial W}{\partial \mathbf C} \mathbf F^\text{T} \right) \right|_{\mathbf F = \mathbf I}  :\mathbf A\\
=\mathbf A \vec \tau +\vec \tau \mathbf A^\text{T} - \vec \tau \tr \mathbf A + 4 \ini{\frac{\partial^{2}  W}{\partial \mathbf C^{2}}}:\mathbf A,
\label{eqn:StressEnergySymmetry}
\end{multline}
for every second-order tensor $\mathbf A$, where we have exploited the fact that $2 \partial W/\partial \mathbf C|_{\mathbf{F}=\mathbf{I}} = \vec \tau$ from equation~\eqref{eqn:StressCompatibility}. We now introduce the linear strain and rotation tensors:
\begin{equation}
 \vec{\varepsilon}=\frac{1}{2}(\nabla\mathbf{u}+(\nabla\mathbf{u})^\text{T})\qquad\text{and}\qquad\vec{\omega}=\frac{1}{2}(\nabla\mathbf{u}-(\nabla\mathbf{u})^\text{T}),
 \label{eqn:LinearStrainTensors}
\end{equation}
respectively, which satisfy $\nabla\mathbf{u}=\vec\varepsilon+\vec\omega$. Substituting $\vec\omega$ for $\mathbf{A}$ in equation~\eqref{eqn:StressEnergySymmetry}, we obtain
\be
\ini{\frac{\partial {\vec \sigma} }{\partial \mathbf F}} :\vec\omega = \vec\omega \vec \tau -\vec \tau \vec \omega,
\label{eqn:OmegaIdentity}
\en
since $\tr\vec\omega=0$ and
\begin{equation}
 \left(\ini{\frac{\partial^2{W}}{\partial^2\mathbf{C}}}:\vec\omega\right)_{ij}=\ini{\frac{\partial^2{W}}{\partial\text{C}_{ji}\partial\text{C}_{\alpha\beta}}}\omega_{\alpha\beta}=-\ini{\frac{\partial^2{W}}{\partial\text{C}_{ji}\partial\text{C}_{\beta\alpha}}}\omega_{\beta\alpha}~\Rightarrow~\ini{\frac{\partial^2{W}}{\partial^2\mathbf{C}}}:\vec\omega=\mathbf{0},
\end{equation}
where we have used the fact that $\vec\omega^\text{T}=-\vec\omega$ and $\mathbf{C}^\text{T}=\mathbf{C}$. Using equations~\eqref{eqn:LinearStrainTensors} and~\eqref{eqn:OmegaIdentity} we can now rewrite equation~\eqref{eqn:LineariseStress} as
\be
	\delta\vec \sigma = \vec \tau + \vec\omega \vec \tau - \vec \tau \vec\omega + \ini{\frac{\partial {\vec  \sigma}}{\partial \mathbf F}}: \vec \varepsilon+\mathcal O((\nabla\mathbf u)^2).
	\label{eqn:LineariseStress2}
\en
At this point, we do not yet know the form of $\partial {\vec \sigma}/\partial \mathbf F|_{\mathbf{F}=\mathbf{I}} :\vec \varepsilon$ explicitly. It could be calculated directly from equation~\eqref{sigmacomp}; however, an alternative approach is to write it as a general rank two symmetric tensor in terms of $\vec \tau$ that is expanded up to first order in $\vec\varepsilon$:
\begin{multline}
	\ini{\frac{\partial  {\vec \sigma}}{\partial \mathbf F}}: \vec \varepsilon = {\alpha_1} \vec \varepsilon+  ({\alpha_2}{\mathbf I} + {\alpha_3}\vec \tau + \alpha_4 \vec \tau^2 )\tr (\vec \varepsilon)  +(  \alpha_5 {\mathbf I} + {\alpha_6} \vec \tau + {\alpha_7} \vec \tau^2) \tr(\vec \varepsilon \vec \tau )
	 \\
	  {\alpha_8} (\vec \varepsilon \vec \tau +\vec \tau \vec \varepsilon) + {\alpha_9} (\vec \varepsilon \vec \tau^2 +\vec \tau^2 \vec \varepsilon)+\mathcal{O}((\nabla\mathbf{u})^2),
	 \label{eqn:CauchyStressLinear}
\end{multline}
where $\alpha_i,~(i=1,...,9)$ are, in general, functions of $I_{\tau_1}$, $I_{\tau_2}$ and $I_{\tau_3}$. Note that neither $\tr(\vec \varepsilon \vec \tau^2)$, $\vec \tau \vec \varepsilon \vec \tau$, $\vec \tau^2 \vec \varepsilon \vec \tau + \vec \tau \vec \varepsilon \vec \tau^2$, nor any power of $\vec \tau$ higher than two is present because they can be written as combinations of the terms already included (see Appendix~\ref{app:Tensors}). For more details on linearising elasticity see \cite{hoger_residual_1993,destrade_third-_2010,destrade_third-_2010-1,shams_initial_2011}.

We now seek to restrict the parameters $\alpha_1,...,\alpha_9$. We begin by rearranging equation~\eqref{eqn:StressEnergySymmetry} and contracting it twice on the left with an arbitrary second-order tensor $\mathbf{B}$, to obtain
\begin{equation}
	4 \mathbf{B}:\ini{\frac{\partial^{2}  W}{\partial \mathbf C^{2}}}:  \mathbf{A}
	=  (\mathbf{B}: \vec \tau) \tr \mathbf{A}  -\mathbf{B}: (\mathbf{A}  \vec \tau) -\mathbf{B}:(\vec \tau\mathbf{A}^\text{T} )   + \mathbf{B}:\ini{\frac{\partial {\vec \sigma}}{\partial \mathbf F}}:\mathbf{A}.
\label{eqn:StressEnergySymmetry2}
\end{equation}
Since equation~\eqref{eqn:StressEnergySymmetry2} must hold for any $\mathbf{A}$ and $\mathbf{B}$, we can swap them to obtain
\begin{equation}
	4 \mathbf{A}:\ini{\frac{\partial^{2}  W}{\partial \mathbf C^{2}}}:  \mathbf{B}
	=  \mathbf{A}: \vec \tau \tr \mathbf{B}  -\mathbf{A}: (\mathbf{B}  \vec \tau) -\mathbf{A}:(\vec \tau\mathbf{B}^\text{T} )   + \mathbf{A}:\ini{\frac{\partial {\vec \sigma}}{\partial \mathbf F}}:\mathbf{B}.
	\label{eqn:StressEnergySymmetry3}
\end{equation}
Now, due to the fact that
\begin{equation}
 \left(\ini{\frac{\partial^2{W}}{\partial^2\mathbf{C}}}\right)_{ijkl}=\left(\ini{\frac{\partial^2{W}}{\partial^2\mathbf{C}}}\right)_{klij}
\end{equation}
we must have
\begin{equation}
\mathbf{A}:\ini{\frac{\partial^{2}  W}{\partial \mathbf C^{2}}}:  \mathbf{B}=\mathbf{B}:\ini{\frac{\partial^{2}  W}{\partial \mathbf C^{2}}}:  \mathbf{A},
\label{eqn:ACB}
\end{equation}
for every $\mathbf{A}$ and $\mathbf{B}$. Upon substituting equations~\eqref{eqn:StressEnergySymmetry2} and~\eqref{eqn:StressEnergySymmetry3} into equation~\eqref{eqn:ACB}, and assuming that $\mathbf{A}$ and $\mathbf{B}$ are small and symmetric, so that equation~\eqref{eqn:CauchyStressLinear} holds with $\mathbf{A}$ and $\mathbf{B}$ substituted for $\vec\varepsilon$, we find that equation~\eqref{eqn:ACB} can hold if and only if
\be
\alpha_{4} =\alpha_{7} =0 \quad  \text{and} \quad \alpha_{5} = \alpha_{3} +1.
\label{eqns:RestrictAlphas}
\en
Substituing the above into equation~\eqref{eqn:LineariseStress2}, we obtain a reduced expression for the stress:
\be
   \boxed{
   \begin{aligned}
     \delta\vec \sigma =&\ \vec \tau + \vec\omega \vec \tau - \vec \tau \vec\omega + \mathbf I \tr(\vec \varepsilon \vec \tau) + {\alpha_1} \vec \varepsilon+  {\alpha_2}\mathbf I\tr (\vec \varepsilon)+  \alpha_{3} \left( \vec \tau \tr (\vec \varepsilon) + {\mathbf I}\tr(\vec \varepsilon \vec \tau ) \right)
	 \\
	  &+  {\alpha_6} \vec \tau \tr(\vec \varepsilon \vec \tau) + {\alpha_8} (\vec \varepsilon \vec \tau +\vec \tau \vec \varepsilon) + {\alpha_9} (\vec \varepsilon \vec \tau^2 +\vec \tau^2 \vec \varepsilon).
   \end{aligned}
   }
	 \label{eqn:LineariseStress3}
\en
In Section~\ref{sec:GeneralIncremental}, we discuss the linearised version of ISRI and its relationship to the linear stress tensor given in equation~\eqref{eqn:LineariseStress3}. When the initial stress is small, we are able to derive a closed-form of the linear stress that satisfies ISRI, as shown in Section~\ref{sec:ModerateInitialStress}.

\subsubsection{Initially stressed neo-Hookean models}
\label{sec:IncNeoHook}

As an aside, we note that if the stress tensors for the initially stressed neo-Hookean models given in equations~\eqref{incNH1} and~\eqref{incNH2} are expanded for small deformations, the resulting linear stress tensors have the above form with
\begin{equation}
 \alpha_1=\frac{2}{K_1}(\mu-\lambda\log K_1),~\alpha_2=\frac{\lambda}{K_1},~\alpha_3=-\alpha_8=-1,~\alpha_6=\alpha_9=0,
\label{eqn:IncNH1}
\end{equation}
for the first model, and
\begin{equation}
 \frac{\alpha_1}{2}=\frac{\mu}{K_2}+\lambda(1-K_2),~\alpha_2=\lambda(2K_2-1),~\alpha_3=-\alpha_8=-1,~\alpha_6=\alpha_9=0,
\label{eqn:IncNH2}
\end{equation}
for the second.

\subsection{The linearised equations of ISRI}
\label{sec:GeneralIncremental}

We now wish to consider the restrictions that are imposed by ISRI in the case of small deformations. We begin by differentiating equation~\eqref{eqn:ISO} with respect to $\xoverline{\mathbf{F}}$ to obtain
\be
\frac{\partial W}{\partial\mathbf{F}}(\widehat {\mathbf F} \xoverline {\mathbf F},\vec \tau )\widehat{\mathbf{F}} = \frac{\partial \xoverline J}{\partial \xoverline {\mathbf F}}W(\widehat {\mathbf F}, \vec \sigma( \xoverline {\mathbf F},\vec \tau) ) + \xoverline J\frac{\partial W}{\partial\vec\sigma}(\widehat {\mathbf F}, \vec \sigma( \xoverline {\mathbf F},\vec \tau) ) \frac{\partial\vec\sigma}{\partial\mathbf{F}}(\xoverline {\mathbf F},\vec \tau ),
\label{eqn:IncISRI}
\en
where $\partial/\partial \mathbf F$ denotes partial differentiation with respect to the first argument of the function and $\partial/\partial \vec \sigma$ denotes partial differentiation with respect to the second. Evaluating equation~\eqref{eqn:IncISRI} at $\widehat {\mathbf F} = \xoverline {\mathbf F} = \mathbf I$ and contracting twice on the right with the linear strain tensor $\vec\varepsilon$ gives
\be
	 \boxed{
	 \vec \tau :\vec \varepsilon =\ \tr \vec \varepsilon \; \ini W+  \ini{\frac{\partial W}{\partial \vec \tau}} : \ini{\frac{\partial   {\vec \sigma}}{\partial \mathbf F}}:\vec \varepsilon \quad \text{for every} \;\; \vec \tau \;\; \text{and} \;\; \vec \varepsilon,
	 }
	 \label{eqn:LinearISO}
\en
which was simplified using equation~\eqref{eqn:StressCompatibility}. One of the terms on the right side can be expanded using the chain rule as follows
\be
	\ini{\frac{\partial W}{\partial \vec \tau}} =	\beta_1\mathbf I +\beta_2\vec \tau +\beta_3\vec \tau^{2},
\label{eqn:DtauW}
\en
where
\be
	\beta_1=\ini{\frac{\partial W}{\partial\tr\vec\tau}}=\ini{\frac{\partial W}{\partial I_{\vec\tau_1}}}+I_{\vec\tau_1}\ini{\frac{\partial W}{\partial I_{\vec\tau_2}}}+I_{\vec\tau_2}\ini{\frac{\partial W}{\partial I_{\vec\tau_3}}}+\ini{\frac{\partial W}{\partial J_1}}+\ini{\frac{\partial W}{\partial J_2}},
\label{eqn:beta1}
\en
	\be
	\beta_2=2\ini{\frac{\partial W}{\partial\tr(\vec\tau)}}=-\ini{\frac{\partial W}{\partial I_{\vec\tau_2}}}-I_{\vec\tau_1}\ini{\frac{\partial W}{\partial I_{\vec\tau_3}}}+2\ini{\frac{\partial W}{\partial J_3}}+2\ini{\frac{\partial W}{\partial J_4}},
\en
\begin{equation}
 \beta_3=3\ini{\frac{\partial W}{\partial\tr(\vec\tau^3)}}=\ini{\frac{\partial W}{\partial I_{\vec\tau_3}}}.
\label{eqn:beta3}
\end{equation}

Using equations~\eqref{eqn:LineariseStress3} and~\eqref{eqn:DtauW} and the Cayley-Hamilton theorem (see Appendix~\ref{app:Tensors}) we can rewrite the restriction~\eqref{eqn:LinearISO} in the form
\be
\tr (\vec \varepsilon\vec \tau) = (\gamma_0 +\ini W) \tr \vec \varepsilon + \gamma_1 \tr(\vec \varepsilon \vec \tau) + \gamma_2 \tr(\vec \varepsilon \vec \tau^2) \quad \text{for every} \;\; \vec \tau \;\; \text{and} \;\; \vec \varepsilon,
\label{eqn:betas}
\en
where $\gamma_0$, $\gamma_1$ and $\gamma_2$ are functions of $\alpha_1,...,\alpha_9$, $\beta_1$, $\beta_2$, $\beta_3$, $I_{\vec\tau_1}$, $I_{\vec\tau_2}$ and $I_{\vec\tau_3}$. Since equation~\eqref{eqn:betas} has to hold for every $\vec \tau$ and $\vec \varepsilon$ (for more details see the supplementary material of~\cite{gower_initial_2015}), we obtain the three equations
\be
\gamma_0 = - \ini W, \quad \gamma_1 =1 \quad \text{and} \quad \gamma_2 =0,
\label{eqns:betas}
\en
which can be written in matrix form as:
\be
	\mathbf M \cdot
	\begin{pmatrix}
	\beta_1
	\\
	\beta_2
	\\
	\beta_3
	\end{pmatrix}
	=
	\begin{pmatrix}
	-\ini W
	\\
	1
	\\
	0
	\end{pmatrix}
	,
	\label{eqn:MWs}
\en
where the matrix $\mathbf M$ depends only on $\alpha_1,...,\alpha_9$, $I_{\vec\tau_1}$, $I_{\vec\tau_2}$ and $I_{\vec\tau_3}$ (the entries of $\mathbf{M}$ are given explicitly in Appendix \ref{App:M}). Since $\beta_1$, $\beta_2$ and $\beta_3$ depend on $\ini W$, the above gives three linear partial differential equations for the single variable $\ini W$. This implies that if $\alpha_1,...\alpha_9$ are unrestricted, $\ini W$ is over-prescribed. Hence, the only way to satisfy equation equation~\eqref{eqn:MWs} is to restrict $\alpha_1,...,\alpha_9$, as we show in the following section.

\subsection{The case of small initial stress}
\label{sec:ModerateInitialStress}

In this section, we assume that the initial stress $\vec\tau$ is small. Our approach is to take the equations~\eqref{eqns:betas} and expand them in powers of $\vec\tau$, neglecting $\mathcal{O}(\|\vec\tau \|^3)$ terms, where $\|\cdot \|$ can be the Frobenius norm or any other equivalent norm. With reference to equation~\eqref{eqn:betas}, we note that $\gamma_1$ multiplies an $\mathcal{O}(\|\vec\tau\|)$ term and $\gamma_2$ multiplies an $\mathcal{O}(\|\vec\tau\|^2)$ term. Therefore, it is only necessary to expand $\gamma_1$ up to $\mathcal{O}(\|\vec\tau\|)$ and $\gamma_2$ up to $\mathcal{O}(\|\vec\tau\|^0)$. Upon doing so, we obtain
\begin{equation}
	\beta_1\left(\alpha_1+3 \alpha_2 + \alpha_3 \tr\vec\tau \right) +\beta_2\left(\alpha_2 \tr\vec\tau + \alpha_3\tr(\vec\tau^2)\right)+\beta_3\alpha_{2}\tr(\vec\tau^2)= - \ini W,
	\label{eqn:beta1b}
\end{equation}
\begin{equation}
	\beta_1\left( 3 (\alpha_3 +1)  + \alpha_6 \tr\vec\tau +2 \alpha_8 \right) + \beta_2\left(\alpha_1 + (\alpha_3+1) \tr\vec\tau \right)  =1,
	\label{eqn:beta2}
\end{equation}
\begin{equation}
	2\beta_1 \alpha_9 + 2\beta_2\alpha_8  + \beta_3\alpha_{1} =0.
	\label{eqn:beta3}
\end{equation}
Next, we expand $\alpha_1,...,\alpha_9$ up to $\mathcal O (\|\vec \tau\|^2)$:
\be
	\alpha_i = \alpha_{i0} +\alpha_{i1} \tr\vec\tau +\alpha_{i2} (\tr\vec\tau)^2 + \alpha_{i3}\tr(\vec\tau^2) \quad \text{for} \quad i=1,2,...,9,
	\label{eqns:alphaSecondOrder}
\en
where the $\alpha_{ij},~i=1,...9,~j=0,...,3$ are constants. For more details on linearising in terms of isotropic invariants see~\cite{destrade_third-_2010}. We also expand $\ini W$ up to $\mathcal O (\|\vec \tau\|^3)$:
\begin{multline}
	\ini W = \psi_0+\psi_1 \tr \vec \tau + \psi_2 (\tr \vec \tau)^2 + \psi_3 \tr( \vec \tau^2) + \psi_4 (\tr \vec \tau )^3 \\
+ \psi_5 \tr \vec \tau \tr(\vec \tau^2)+ \psi_6 \tr( \vec \tau^3),
	\label{eqns:WtauSecondOrder}
\end{multline}
where $\psi_0,...,\psi_6$ are constants and we immediately choose $\psi_0=0$ since we expect
\begin{equation}
\lim_{\vec\tau\rightarrow\mathbf{0}}\ini W=0.
\end{equation}
Upon substituting equation~\eqref{eqns:WtauSecondOrder} into equations~\eqref{eqn:beta1}--\eqref{eqn:beta3}, we obtain $\beta_1$, $\beta_2$ and $\beta_3$ expanded up to $\mathcal O(\|\vec \tau\|^2)$, $\mathcal O(\|\vec \tau\|^1)$ and $\mathcal O(\|\vec \tau\|^0)$, respectively, which can then be substituted into equations~\eqref{eqn:beta1b}--\eqref{eqn:beta3}. We then solve the resulting system of equations for the parameters $\alpha_{ij}$ and $\psi_i$, where we note that the stress tensor of an initially stressed material must generalise that derived from classical linear elasticity. In other words, when $\vec \tau \to \vec 0$ we must have
 \be
\delta\vec\sigma= \alpha_{10}\vec\varepsilon + \alpha_{20}{\mathbf I} \tr (\vec \varepsilon),\quad\text{where}\quad\alpha_{10}=2\mu\quad\text{and}\quad\alpha_{20}=\lambda,
	\label{eqn:linearConstitency}
\en
where $\lambda$ and $\mu$ are the first and second Lam\'e parameters, respectively. Using equation~\eqref{eqn:linearConstitency}, the final system of equations simplifies to the following conditions:
\bal
	 &\psi_1=0,\qquad\psi_2=-\frac{\lambda}{12\kappa\mu},\qquad\psi_3=\frac{1}{4\mu},
	\\
	 &\psi_4=\frac{2\lambda^2(3\alpha_{11}-2\alpha_{80})+2\lambda\mu(4\alpha_{11}+4\alpha_{30}+3)-8\mu^2\alpha_{21}}{216\kappa^2\mu^2},
	\\
	 &\psi_5=\frac{\lambda(2\alpha_{80}-3\alpha_{11})-2\mu(\alpha_{11}+\alpha_{30}+1)}{24\kappa\mu^2},\qquad\psi_6=-\frac{\alpha_{80}}{6\mu^2},
	\\
	 &\alpha_{80}=\frac{2\mu\alpha_{30}-3\kappa\alpha_{11}}{2\lambda},
	\label{eqns:Alpha8Restriction}
\eal
where $\kappa=\lambda+2\mu/3$ is the bulk modulus of the material under consideration. Equation~\eqref{eqns:Alpha8Restriction} relates $\alpha_{80}$ to $\lambda$, $\mu$, $\alpha_{11}$ and $\alpha_{30}$, and therefore reduces the number of free parameters in the system by one. We now use the above to write the linearised Cauchy stress in terms of the strain and initial stress:
\be
\boxed{
\begin{aligned}
\delta\vec\sigma=&\ \vec\tau+\vec\omega\vec\tau-\vec\tau\vec\omega+\mathbf I\tr(\vec\varepsilon\vec\tau)+2(\mu+\mu_1\tr\vec\tau)\vec\varepsilon+(\lambda+\lambda_1\tr\vec\tau)\mathbf I\tr(\vec\varepsilon)\\
&+\eta\left(\vec\tau\tr(\vec\varepsilon)+\mathbf I\tr(\vec\varepsilon\vec\tau)\right)+\left(\frac{\mu\eta}{\lambda}-\frac{3\kappa\mu_1}{2\lambda}\right)(\vec\varepsilon\vec\tau+\vec\tau\vec\varepsilon),
\end{aligned}}
\label{eqn:LineariseStressAndTau}
\en
where we have renamed $\alpha_{11} = 2\mu_{1}$, ${\alpha_{21}} = \lambda_{1}$ and $\alpha_{30} = \eta$ and all the parameters in the equation above are constants. Equation~\eqref{eqn:LineariseStressAndTau} differs from the stress tensor first deduced in~\cite{man_hartigs_1998} because of the restriction given in equation~\eqref{eqns:Alpha8Restriction}. The parameters above may be further restricted by considerations such as strong-ellipticity~\cite{walton_sufficient_2003,han_conditions_2009}, but ultimately, they can be determined by ultrasonic, indentation, or hole drilling experiments.

\subsubsection{Initially stressed neo-Hookean models}

If equations~\eqref{eqn:K1} and~\eqref{eqn:K2} are expanded for small $\vec\tau$, they can be solved for $K_1$ and $K_2$, which have the same series expansion up to order one in $\vec\tau$:
\begin{equation}
 K_1=K_2=1+\frac{I_{\vec\tau_1}}{3\kappa}+\mathcal{O}(\vec\tau^2).
\label{eqn:K1K2}
\end{equation}
Equation~\eqref{eqn:K1K2} can then be substituted into equations~\eqref{eqn:IncNH1} and~\eqref{eqn:IncNH2} to obtain
\begin{equation}
 \alpha_1=2\mu-\frac{2(\lambda+\mu)}{3\kappa}I_{\vec\tau_1}+\mathcal{O}(\vec\tau^2),\qquad\alpha_2=\lambda-\frac{\lambda}{3\kappa}I_{\vec\tau_1}+\mathcal{O}(\vec\tau^2),
\end{equation}
for the first model, and
\begin{equation}
 \alpha_1=2\mu-\frac{2(\lambda+\mu)}{3\kappa}I_{\vec\tau_1}+\mathcal{O}(\vec\tau^2),\qquad\alpha_2=\lambda+\frac{2\lambda}{3\kappa}I_{\vec\tau_1}+\mathcal{O}(\vec\tau^2).
\end{equation}
for the second. Therefore, for both models, we have
\begin{equation}
 \alpha_{10}=2\mu,\quad\alpha_{11}=-\frac{2(\lambda+\mu)}{3\kappa},\quad\alpha_{20}=\lambda\quad\alpha_{30}=-1,\quad\text{and}\quad\alpha_{80}=1,
\end{equation}
which satisfy equation~\eqref{eqns:Alpha8Restriction}, as required. The linearised stress tensors associated with the two models are
\bal
	\delta\vec\sigma_\text{GSC1}= & \ \vec\tau+\vec\omega\vec\tau-\vec\tau\vec\omega-\vec\tau\tr(\vec \varepsilon)+2\left(\mu-\frac{\lambda+\mu}{3\kappa}\tr\vec\tau\right)\vec\varepsilon{}
	\\
	&+\left(\lambda-\frac{\lambda}{3\kappa}\tr\vec\tau\right)\mathbf I\tr(\vec\varepsilon)+\vec\varepsilon\vec\tau+\vec \tau\vec\varepsilon,
\eal
and
\bal
	\delta\vec\sigma_\text{GSC2}= &\ \vec\tau+\vec\omega\vec\tau-\vec\tau\vec\omega-\vec\tau\tr(\vec \varepsilon)+2\left(\mu-\frac{\lambda+\mu}{3\kappa}\tr\vec\tau\right)\vec\varepsilon{}
	\\
	& +\left(\lambda+\frac{2\lambda}{3\kappa}\tr\vec\tau\right)\mathbf I\tr(\vec\varepsilon)+\vec\varepsilon\vec\tau+\vec \tau\vec\varepsilon.
\eal

\section{Discussion}
\label{sec:Discussion}
Most constitutive choices in the literature of the form $W := W(\mathbf F, \vec \tau)$ do not satisfy the ISRI restrictions~\eqref{eqn:ISO} and~\eqref{eqn:LinearISO} presented in this paper. In Section~\ref{sec:uniaxial} we gave an example of how these constitutive choices may lead to unphysical behaviour even for simple deformations such as uniaxial extension. This is also true of more complex deformations. Taking an example from biomechanics, where residual stresses play a crucial role, suppose we wish to model the mechanics of an arterial wall that supports an internal pressure. Let us choose two different reference configurations: first, the \emph{unloaded configuration} where the fluid in the artery has been removed, and second, the \emph{opening angle configuration}~\cite{rodriguez_stress-dependent_1994,johnson_use_1995} where the fluid has been removed and the artery has been cut along its axis.  Both these configurations are subject to no external loads, but there will be less (and differently distributed) internal stress in the opening angle configuration. If we use a strain energy function $W(\mathbf F, \vec \tau)$ that does not satisfy ISRI, then each of the two reference configurations will lead to a different stress distribution in the intact, inflated configuration of the arterial wall. We therefore cannot believe the preditions from either reference configuration since a physically correct model should \emph{not} give different results due to an arbitrary choice of reference configuration.

By using ISRI we were able to derive a restricted form for the linear elastic stress tensor~\eqref{eqn:LineariseStressAndTau} in the case of small initial stress. This reduced form may ultimately improve material characterisation based on ultrasonic and indentation experiments. Many studies (see \cite{man_hartigs_1998} and the references therein) have confirmed that a linearised stress tensor of the form given in equation~\eqref{eqn:LineariseStressAndTau} is well suited to fitting experimental data.

One outstanding problem for metals~\cite{totten_handbook_2002}, biological soft tissues and other materials~\cite{man_separation_1996} is the difficulty in differentiating between the effects of structural anisotropy~\cite{spencer_deformations_1972} and anisotropy caused by initial stress. The linear form of ISRI given in equation~\eqref{eqn:LinearISO} will help to differentiate between these effects, as it dictates a specific dependency of the elastic stress on the initial stress. Nevertheless, future work should focus on developing the consequences of ISRI for materials with structural anisotropy. This will be particularly important for collagenous soft tissues, which are known to be structurally anisotropic due to the presence of collagen fibres~\cite{shearer2015new,shearer2015new2}. Initial stresses in soft tissues can be significant~\cite{rodriguez_stress-dependent_1994,joshi_reconstruction_2013,ciarletta_morphology_2016}, so assuming a small initial stress may not give accurate predictions. Currently, the internal stress in soft tissues is often measured by excising a sample and then estimating its initial deformation from a theoretically stress-free configuration. To measure stress in-vivo, non-invasive techniques need to be improved. Ultrasound techniques are among the most suitable and promising methods for measuring initial stress~\cite{chen_youngs_1996,han_novel_2003}, and the ISRI restrictions could ultimately improve them.




\section*{Acknowledgements}
A.L.G. gratefully acknowledges funding provided by the EPSRC (no. EP/M026205/1). T.S. would also like the thank the EPSRC for supporting this work via his Fellowship grant (no. EP/L017997/1). P.C acknowledges funding from the AIRC grant MFAG 17412. The authors would also like to thank Professor M. Destrade for his feedback and Professor R. Ogden for helpful discussions.

\bibliographystyle{plain}
\bibliography{Elasticity}

\begin{thebibliography}{10}

\bibitem{alessandrini_inverse_2003}
Giovanni Alessandrini and Gunther Uhlmann.
\newblock {\em Inverse {Problems}: {Theory} and {Applications} : {INdAM}
  {Workshop} on {Inverse} {Problems} and {Applications}, {June} 3-9, 2002,
  {Cortona}, {Italy} : {Special} {Session} at {AMS}-{UMI} {First} {Joint}
  {International} {Meeting} on {Inverse} {Boundary} {Problems} and
  {Applications}, {June} 12-16, 2002, {Pisa}, {Italy}}.
\newblock American Mathematical Soc., 2003.

\bibitem{bustamante2011solutions}
R~Bustamante and KR~Rajagopal.
\newblock Solutions of some simple boundary value problems within the context
  of a new class of elastic materials.
\newblock {\em International Journal of Non-Linear Mechanics}, 46(2):376--386,
  2011.

\bibitem{castellano_monitoring_2016}
A.~Castellano, P.~Foti, A.~Fraddosio, S.~Marzano, F.~Paparella, and M.~Daniele
  Piccioni.
\newblock Monitoring applied and residual stress in materials and structures by
  non-destructive acoustoelastic techniques.
\newblock In {\em 2016 {IEEE} {Workshop} on {Environmental}, {Energy}, and
  {Structural} {Monitoring} {Systems} ({EESMS})}, pages 1--5, June 2016.

\bibitem{chen_youngs_1996}
E.~J. Chen, J.~Novakofski, W.~K. Jenkins, and W.~D. O'Brien.
\newblock Young's modulus measurements of soft tissues with application to
  elasticity imaging.
\newblock {\em IEEE Transactions on Ultrasonics, Ferroelectrics, and Frequency
  Control}, 43(1):191--194, January 1996.

\bibitem{chuong_residual_1986}
CJ~Chuong and YC~Fung.
\newblock Residual stress in arteries.
\newblock In {\em Frontiers in {Biomechanics}}, pages 117--129. Springer, 1986.

\bibitem{ciarletta_morphology_2016}
P.~Ciarletta, M.~Destrade, A.~L. Gower, and M.~Taffetani.
\newblock Morphology of residually stressed tubular tissues: {Beyond} the
  elastic multiplicative decomposition.
\newblock {\em Journal of the Mechanics and Physics of Solids}, 90:242--253,
  May 2016.

\bibitem{destrade_third-_2010}
Michel Destrade, Michael~D. Gilchrist, and Raymond~W. Ogden.
\newblock Third- and fourth-order elasticities of biological soft tissues.
\newblock {\em The Journal of the Acoustical Society of America}, 127(4):2103,
  2010.

\bibitem{destrade_third-_2010-1}
Michel Destrade, Michael~D. Gilchrist, and Giuseppe Saccomandi.
\newblock Third- and fourth-order constants of incompressible soft solids and
  the acousto-elastic effect.
\newblock {\em The Journal of the Acoustical Society of America}, 127(5):2759,
  2010.

\bibitem{fung_what_1991}
YC~Fung.
\newblock What are the residual stresses doing in our blood vessels?
\newblock {\em Annals of biomedical engineering}, 19(3):237--249, 1991.

\bibitem{gandhi_acoustoelastic_2012}
Navneet Gandhi, Jennifer~E. Michaels, and Sang~Jun Lee.
\newblock Acoustoelastic {Lamb} wave propagation in biaxially stressed plates.
\newblock {\em The Journal of the Acoustical Society of America},
  132(3):1284--1293, September 2012.

\bibitem{gower_initial_2015}
A.~L. Gower, P.~Ciarletta, and M.~Destrade.
\newblock Initial stress symmetry and its applications in elasticity.
\newblock {\em Proceedings of the Royal Society of London A: Mathematical,
  Physical and Engineering Sciences}, 471(2183), 2015.

\bibitem{guillou_growth_2006}
A.~Guillou and R.~W. Ogden.
\newblock {\em Growth in soft biological tissue and residual stress
  development}.
\newblock Springer, 2006.

\bibitem{guz2002elastic}
AN~Guz.
\newblock Elastic waves in bodies with initial (residual) stresses.
\newblock {\em International Applied Mechanics}, 38(1):23--59, 2002.

\bibitem{han_conditions_2009}
Deren Han, H.~H. Dai, and Liqun Qi.
\newblock Conditions for {Strong} {Ellipticity} of {Anisotropic} {Elastic}
  {Materials}.
\newblock {\em J Elasticity}, 97(1):1--13, May 2009.

\bibitem{han_novel_2003}
Lianghao Han, J.~Alison Noble, and Michael Burcher.
\newblock A novel ultrasound indentation system for measuring biomechanical
  properties of in vivo soft tissue.
\newblock {\em Ultrasound in Medicine \& Biology}, 29(6):813--823, June 2003.

\bibitem{hoger_determination_1986}
Anne Hoger.
\newblock On the determination of residual stress in an elastic body.
\newblock {\em Journal of Elasticity}, 16(3):303--324, 1986.

\bibitem{hoger_elasticity_1993}
Anne Hoger.
\newblock The elasticity tensors of a residually stressed material.
\newblock {\em Journal of elasticity}, 31(3):219--237, 1993.

\bibitem{hoger_residual_1993}
Anne Hoger.
\newblock Residual stress in an elastic body: a theory for small strains and
  arbitrary rotations.
\newblock {\em J Elasticity}, 31(1):1--24, April 1993.

\bibitem{holzapfel_biomechanics_2003}
Gerhard~A Holzapfel and Ray~W Ogden.
\newblock {\em Biomechanics of soft tissue in cardiovascular systems}, volume
  441.
\newblock Springer Science \& Business Media, 2003.

\bibitem{james2010shot}
MN~James, M~Newby, DG~Hattingh, and A~Steuwer.
\newblock Shot-peening of steam turbine blades: Residual stresses and their
  modification by fatigue cycling.
\newblock {\em Procedia Engineering}, 2(1):441--451, 2010.

\bibitem{johnson_use_1995}
Byron~E Johnson and Anne Hoger.
\newblock The use of a virtual configuration in formulating constitutive
  equations for residually stressed elastic materials.
\newblock {\em Journal of Elasticity}, 41(3):177--215, 1995.

\bibitem{joshi_reconstruction_2013}
Sunnie Joshi and Jay~R. Walton.
\newblock Reconstruction of the residual stresses in a hyperelastic body using
  ultrasound techniques.
\newblock {\em International Journal of Engineering Science}, 70:46--72,
  September 2013.

\bibitem{kubrusly_derivation_2016}
Alan~C. Kubrusly, Arthur M.~B. Braga, and Jean Pierre von~der Weid.
\newblock Derivation of acoustoelastic {Lamb} wave dispersion curves in
  anisotropic plates at the initial and natural frames of reference.
\newblock {\em The Journal of the Acoustical Society of America},
  140(4):2412--2417, October 2016.

\bibitem{lennon_residual_2002}
A.~B. Lennon and P.~J. Prendergast.
\newblock Residual stress due to curing can initiate damage in porous bone
  cement: experimental and theoretical evidence.
\newblock {\em J Biomech}, 35(3):311--321, March 2002.

\bibitem{lin_uniqueness_2003}
Ching-Lung Lin and Jenn-Nan Wang.
\newblock Uniqueness in inverse problems for an elasticity system with residual
  stress by a single measurement.
\newblock {\em Inverse Problems}, 19(4):807, 2003.

\bibitem{lu_covariant_2012}
J.~Lu.
\newblock A covariant constitutive theory for anisotropic hyperelastic solids
  with initial strains.
\newblock {\em Mathematics and Mechanics of Solids}, 17(2):104--119, March
  2012.

\bibitem{man_hartigs_1998}
Chi-Sing Man.
\newblock Hartig's law and linear elasticity with initial stress.
\newblock {\em Inverse Problems}, 14(2):313, 1998.

\bibitem{man_towards_1987}
Chi-Sing Man and WY~Lu.
\newblock Towards an acoustoelastic theory for measurement of residual stress.
\newblock {\em Journal of Elasticity}, 17(2):159--182, 1987.

\bibitem{man_separation_1996}
Chi-Sing Man and Roberto Paroni.
\newblock On the separation of stress-induced and texture-induced birefringence
  in acoustoelasticity.
\newblock {\em J Elasticity}, 45(2):91--116, November 1996.

\bibitem{marsden_mathematical_1994}
Jerrold~E Marsden and Thomas~JR Hughes.
\newblock {\em Mathematical foundations of elasticity}.
\newblock Dover publications, 1994.

\bibitem{merodio_extension_2015}
José Merodio and Ray~W Ogden.
\newblock Extension, inflation and torsion of a residually stressed circular
  cylindrical tube.
\newblock {\em Continuum Mechanics and Thermodynamics}, pages 1--18, 2015.

\bibitem{merodio_influence_2013}
José Merodio, Ray~W Ogden, and Javier Rodríguez.
\newblock The influence of residual stress on finite deformation elastic
  response.
\newblock {\em International Journal of Non-Linear Mechanics}, 56:43--49,
  November 2013.

\bibitem{nam_effect_2016}
N.~T. Nam, J.~Merodio, R.~W. Ogden, and P.~C. Vinh.
\newblock The effect of initial stress on the propagation of surface waves in a
  layered half-space.
\newblock {\em International Journal of Solids and Structures},
  88–89:88--100, June 2016.

\bibitem{ogden_non-linear_1997}
Ray~W Ogden.
\newblock {\em Non-linear elastic deformations}.
\newblock Courier Dover Publications, 1997.

\bibitem{parnell2012employing}
William~J Parnell, Andrew~N Norris, and Tom Shearer.
\newblock Employing pre-stress to generate finite cloaks for antiplane elastic
  waves.
\newblock {\em Applied Physics Letters}, 100(17):171907, 2012.

\bibitem{parnell2013antiplane}
William~J Parnell and Tom Shearer.
\newblock Antiplane elastic wave cloaking using metamaterials, homogenization
  and hyperelasticity.
\newblock {\em Wave Motion}, 50(7):1140--1152, 2013.

\bibitem{rachele_uniqueness_2003}
Lizabeth~V. Rachele.
\newblock Uniqueness in {Inverse} {Problems} for {Elastic} {Media} with
  {Residual} {Stress}.
\newblock {\em Communications in Partial Differential Equations},
  28(11-12):1787--1806, January 2003.

\bibitem{rajagopal_implicit_2003}
K.~R. Rajagopal.
\newblock On {Implicit} {Constitutive} {Theories}.
\newblock {\em Applications of Mathematics}, 48(4):279--319, 2003.

\bibitem{rajagopal_response_2007}
K.~R. Rajagopal and A.~R. Srinivasa.
\newblock On the response of non-dissipative solids.
\newblock {\em Proceedings of the Royal Society of London A: Mathematical,
  Physical and Engineering Sciences}, 463(2078):357--367, February 2007.

\bibitem{robertson_determining_1998}
Robert~L. Robertson.
\newblock Determining {Residual} {Stress} from {Boundary} {Measurements}: {A}
  {Linearized} {Approach}.
\newblock {\em Journal of Elasticity}, 52(1):63--73, July 1998.

\bibitem{rodriguez_stress-dependent_1994}
Edward~K. Rodriguez, Anne Hoger, and Andrew~D. McCulloch.
\newblock Stress-dependent finite growth in soft elastic tissues.
\newblock {\em Journal of Biomechanics}, 27(4):455--467, April 1994.

\bibitem{rossini_methods_2012}
N.~S. Rossini, M.~Dassisti, K.~Y. Benyounis, and A.~G. Olabi.
\newblock Methods of measuring residual stresses in components.
\newblock {\em Materials \& Design}, 35:572--588, March 2012.

\bibitem{shams_effect_2016}
M.~Shams.
\newblock Effect of initial stress on {Love} wave propagation at the boundary
  between a layer and a half-space.
\newblock {\em Wave Motion}, 65:92--104, September 2016.

\bibitem{shams_initial_2011}
M.~Shams, M.~Destrade, and R.~W. Ogden.
\newblock Initial stresses in elastic solids: {Constitutive} laws and
  acoustoelasticity.
\newblock {\em Wave Motion}, 48(7):552 -- 567, 2011.

\bibitem{shams_wave_2010}
Moniba Shams.
\newblock {\em Wave propagation in residually-stressed materials}.
\newblock {PhD}, University of Glasgow, 2010.

\bibitem{shams_rayleigh-type_2014}
Moniba Shams and Ray~W Ogden.
\newblock On {Rayleigh}-type surface waves in an initially stressed
  incompressible elastic solid.
\newblock {\em IMA Journal of Applied Mathematics}, 79(2):360--376, 2014.

\bibitem{sharafutdinov_tomography_2012}
Vladimir Sharafutdinov and Jenn-Nan Wang.
\newblock Tomography of small residual stresses.
\newblock {\em Inverse Problems}, 28(6):065017, 2012.

\bibitem{shearer2015new2}
Tom Shearer.
\newblock A new strain energy function for modelling ligaments and tendons
  whose fascicles have a helical arrangement of fibrils.
\newblock {\em Journal of biomechanics}, 48(12):3017--3025, 2015.

\bibitem{shearer2015new}
Tom Shearer.
\newblock A new strain energy function for the hyperelastic modelling of
  ligaments and tendons based on fascicle microstructure.
\newblock {\em Journal of biomechanics}, 48(2):290--297, 2015.

\bibitem{shearer2013torsional}
Tom Shearer, I~David Abrahams, William~J Parnell, and Carlos~H Daros.
\newblock Torsional wave propagation in a pre-stressed hyperelastic annular
  circular cylinder.
\newblock {\em The Quarterly Journal of Mechanics and Applied Mathematics},
  66(4):465--487, 2013.

\bibitem{shearer2015antiplane}
Tom Shearer, William~J Parnell, and I~David Abrahams.
\newblock Antiplane wave scattering from a cylindrical cavity in pre-stressed
  nonlinear elastic media.
\newblock {\em Proceedings of the Royal Society A}, 471(2182):20150450, 2015.

\bibitem{spencer_theory_1971}
AJM Spencer.
\newblock Theory of invariants.
\newblock In {\em Continuum physics, {Vol}. 1}, volume~1, pages 239--353.
  Academic Press, 1971.

\bibitem{spencer_deformations_1972}
Anthony James~Merrill Spencer.
\newblock {\em Deformations of fibre-reinforced materials}.
\newblock Oxford University Press, 1972.

\bibitem{taber_stress-modulated_2001}
Larry~A. Taber and Jay~D. Humphrey.
\newblock Stress-{Modulated} {Growth}, {Residual} {Stress}, and {Vascular}
  {Heterogeneity}.
\newblock {\em J Biomech Eng}, 123(6):528--535, July 2001.

\bibitem{tanuma_perturbation_2008}
Kazumi Tanuma and Chi-Sing Man.
\newblock Perturbation {Formulas} for {Polarization} {Ratio} and {Phase}
  {Shift} of {Rayleigh} {Waves} in {Prestressed} {Anisotropic} {Media}.
\newblock {\em J Elasticity}, 92(1):1--33, July 2008.

\bibitem{thurston_third-order_1964}
R.~N. Thurston and K.~Brugger.
\newblock Third-{Order} {Elastic} {Constants} and the {Velocity} of {Small}
  {Amplitude} {Elastic} {Waves} in {Homogeneously} {Stressed} {Media}.
\newblock {\em Phys. Rev.}, 133(6A):A1604--A1610, March 1964.

\bibitem{todd1999thermal}
RI~Todd, AR~Boccaccini, R~Sinclair, RB~Yallee, and RJ~Young.
\newblock Thermal residual stresses and their toughening effect in al 2 o 3
  platelet reinforced glass.
\newblock {\em Acta materialia}, 47(11):3233--3240, 1999.

\bibitem{totten_handbook_2002}
George~E. Totten.
\newblock {\em Handbook of {Residual} {Stress} and {Deformation} of {Steel}}.
\newblock ASM International, 2002.

\bibitem{walton_sufficient_2003}
Jay~R. Walton and J.~Patrick Wilber.
\newblock Sufficient conditions for strong ellipticity for a class of
  anisotropic materials.
\newblock {\em International journal of non-linear mechanics}, 38(4):441--455,
  2003.

\bibitem{webster_residual_2001}
G.~A. Webster and A.~N. Ezeilo.
\newblock Residual stress distributions and their influence on fatigue
  lifetimes.
\newblock {\em International Journal of Fatigue}, 23, Supplement 1:375--383,
  2001.

\end{thebibliography}

\appendix

\section{Deduction of the strain energy function $W_\text{GCD}$}
\label{app:DeducingValidW}

\begin{figure}[ht!]
\centering{
\begin{tikzpicture} [scale=0.9]
	\draw (-7,-3) node {\includegraphics[width=0.32\textwidth]{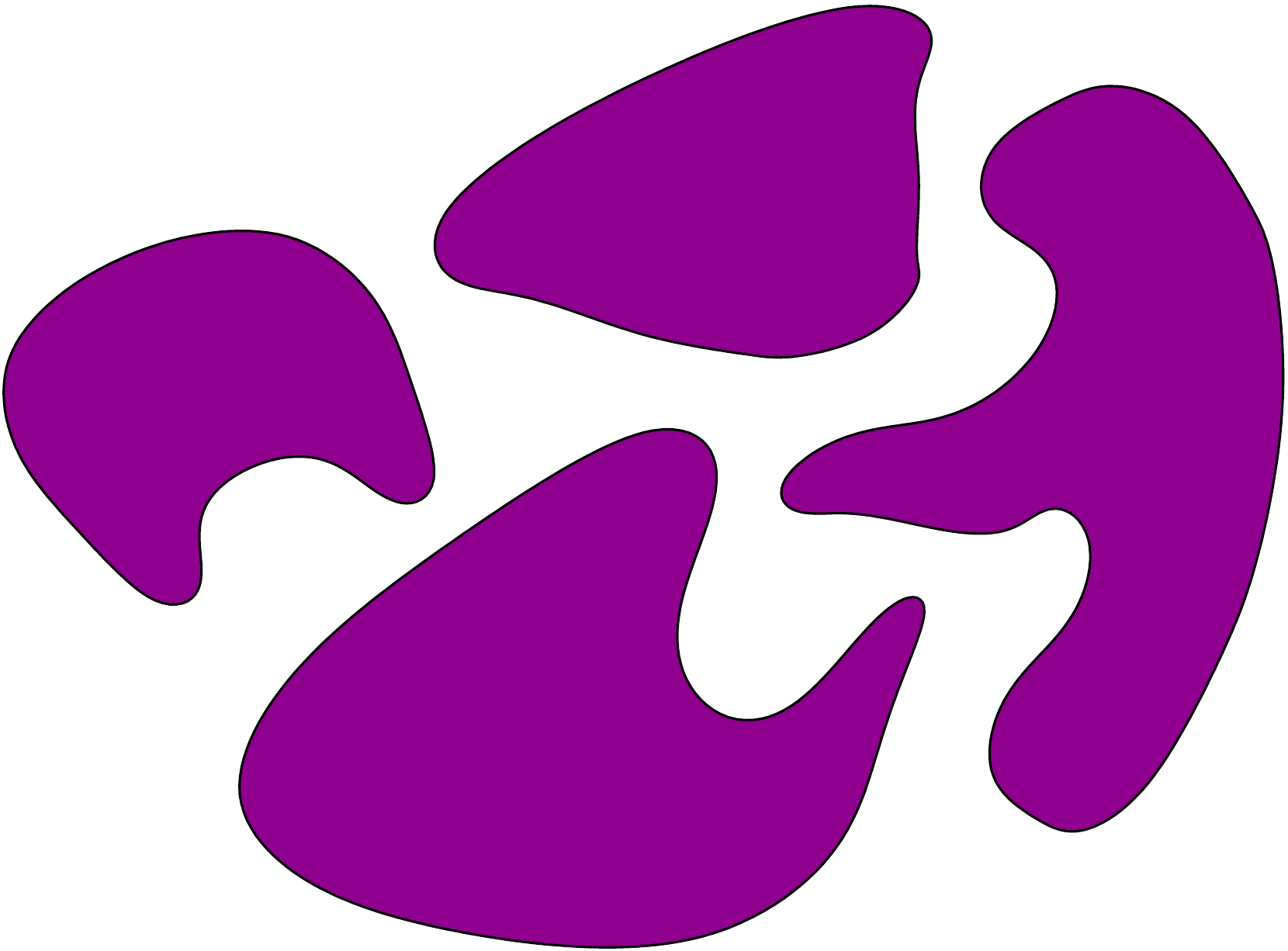}};
 	\draw (3.2,-3.7) node {\includegraphics[width=0.32\textwidth]{blob3.pdf}};
 	\draw (2.8,-2.5) node {$\xoverline{\bb}$};
 	\draw (-1,0) node {\includegraphics[width=0.3\textwidth]{blob1.pdf}};
 	\draw (-2.6,1.6) node {$\bb$} ;
 	\draw (-8,-1.6) node {$\bb_0$} ;
 	\draw (-1,0.1) node{$\boldsymbol{\tau}$};
 	\draw (3.9,-3.6) node{$\xoverline{\boldsymbol{\sigma}}$};
 	\draw [->] (-5.5,-1.5) .. controls (-5.,-0.4) and (-3.6,-0.2) .. (-3.3,-0.2)[thick];
 	\draw (-5,-0.6) node[above]{$\mathbf{F}_0$};
 	\draw [->] (0.1,-1.4) .. controls (1.5,-2.) and (2,-3.) .. (2.1,-3.4)[thick];
 	\draw (1.55,-1.55) node[below]{$\xoverline{\mathbf{F}}$};
 	\draw [->] (-5.0,-4.1) .. controls (-4,-5) and (-1.5,-5.2) .. (0.75,-5.0)[thick];
 	\draw (-2.5,-5.4) node[right]{$\mathbf{F}_1$};
\end{tikzpicture}
}
\caption{Deformation of an incompressible neo-Hookean material from a stress-free configuration $\mathcal{B}_0$ to the stressed configurations $\mathcal{B}$ and $\xoverline{\mathcal{B}}$.}
\label{fig:Configurations2}
\end{figure}

The strain energy function \eqref{WGCD} was first derived in \cite{gower_initial_2015}. Here, an alternative derivation is presented by considering deformations of an incompressible neo-Hookean material from a stress-free configuration $\mathcal{B}_0$ to the stressed configurations $\mathcal{B}$ and $\xoverline{\mathcal{B}}$ (see Figure \ref{fig:Configurations2} and compare with Figure \ref{fig:Configurations}).

The neo-Hookean strain energy function is given by
\begin{equation}
W_\text{NH}=\mu(I_1-3),
\label{WNH}
\end{equation}
where $\mu$ is the ground state shear modulus of the material under consideration. Upon substituting \eqref{WNH} into \eqref{sigmacomp} with $W_{I_3}=0$ (because the material is incompressibile) and then taking $\mathbf{F}=\mathbf{F}_0$ and $\mathbf{F}=\mathbf{F}_1$, it follows that
\begin{equation}
\boldsymbol{\tau}=\mu\mathbf{B}_0-p_0\mathbf{I} \quad \text{and} \quad\xoverline{\boldsymbol{\sigma}}=\mu\mathbf{B}_1-p_1\mathbf{I},
\label{tauandsigmabar}
\end{equation}
where $\mathbf{B}_0=\mathbf{F}_0\mathbf{F}_0^\text{T}$,  $\mathbf{B}_1=\mathbf{F}_1\mathbf{F}_1^\text{T}$ and $p_0$ and $p_1$ are the Lagrange multipliers associated with the two respective deformations. By rearranging equation~\eqref{tauandsigmabar}${}_1$ and taking the determinant of both sides, the following is obtained:
\begin{equation}
\det(\mu\mathbf{B}_0)=\det(\boldsymbol{\tau}+p_0\mathbf{I})~~~~~\Leftrightarrow~~~~~\mu^3=p_0^3+p_0^2I_{\boldsymbol{\tau}_1}+p_0I_{\boldsymbol{\tau}_2}+I_{\boldsymbol{\tau}_3},
\end{equation}
where $\det(\mathbf{B}_0)=1$ because the material is incompressible. Only one of the three roots of the above polynomial is physically meaningful~\cite{gower_initial_2015} and it is given by equation \eqref{eqn:p0}. Using $\mathbf{F}_1=\xoverline{\mathbf{F}}\mathbf{F}_0$, equation \eqref{tauandsigmabar}${}_2$ gives
\begin{equation}
\xoverline{\boldsymbol{\sigma}}=\mu\xoverline{\mathbf{F}}\mathbf{B}_0\xoverline{\mathbf{F}}^\text{T}-p_1\mathbf{I}.
\label{sigmabar}
\end{equation}
The aim is to derive an initially stressed strain energy function that gives equation \eqref{sigmabar} with $\mathcal{B}$ as the reference configuration. For simplicity, it is assumed that the strain energy function depends only upon $I_1$, $J_1$ and the three initial stress invariants $I_{\boldsymbol{\tau}_1}$, $I_{\boldsymbol{\tau}_2}$ and $I_{\boldsymbol{\tau}_3}$. Making this assumption and substituting $\mathbf{F}=\xoverline{\mathbf{F}}$ into equation \eqref{sigmacomp} with $W_{I_3}=0$, it follows that
\begin{align}
\xoverline{\boldsymbol{\sigma}}=\boldsymbol{\sigma}(\xoverline{\mathbf{F}},\boldsymbol{\tau})&=2W_1\xoverline{\mathbf{B}}+2W_{J_1}\xoverline{\mathbf{F}}\boldsymbol{\tau}\xoverline{\mathbf{F}}^\text{T}-\xoverline{p}\mathbf{I}\label{WGCDDev}\\
&=2W_1\xoverline{\mathbf{B}}+2W_{J_1}(\mu\xoverline{\mathbf{F}}\mathbf{B}_0\xoverline{\mathbf{F}}^\text{T}-p_0\xoverline{\mathbf{B}})-\xoverline{p}\mathbf{I}.
\label{WGCDderivation}
\end{align}
For equation \eqref{WGCDderivation} to be equivalent to equation \eqref{sigmabar}, the following equations must be satisfied:
\begin{equation}
 2W_1=p_0,\qquad2W_{J_1}=1,\qquad\xoverline{p}=p_1.
\label{WCDconditions}
\end{equation}
The third of these equations does not tell us anything about the required functional form of $W$; however, upon solving the first two, the following is obtained:
\begin{equation}
W=\frac{1}{2}(p_0(I_{\boldsymbol{\tau}_1},I_{\boldsymbol{\tau}_2},I_{\boldsymbol{\tau}_3})I_1+J_1)+f(I_{\boldsymbol{\tau}_1},I_{\boldsymbol{\tau}_2},I_{\boldsymbol{\tau}_3}),
\label{WCD1}
\end{equation}
where $f$ is an arbitrary function of $I_{\boldsymbol{\tau}_1}$, $I_{\boldsymbol{\tau}_2}$ and $I_{\boldsymbol{\tau}_3}$. Upon choosing $f(I_{\boldsymbol{\tau}_1},I_{\boldsymbol{\tau}_2},I_{\boldsymbol{\tau}_3})=-\frac{3}{2}\mu$, the final form of the strain energy function \eqref{WGCD} is obtained. This choice ensures that the energy derived using the initially stressed strain energy function is the same as that obtained by considering a direct deformation of a neo-Hookean material from the stress-free configuration.

All that remains is to prove that, when using $W_\text{GCD}$, the third equation of \eqref{WCDconditions} holds. Equations \eqref{tauandsigmabar}${}_1$ and \eqref{sigmabar} can be rearranged to give
\begin{equation}
p_0\mathbf{I}=\mu\mathbf{B}_0-\boldsymbol{\tau}~~~~~\text{and}~~~~~p_1\mathbf{I}=\mu\xoverline{\mathbf{F}}\mathbf{B}_0\xoverline{\mathbf{F}}^\text{T}-\xoverline{\boldsymbol{\sigma}},
\label{Dev}
\end{equation}
respectively. Multiplying the first of these equations on the left by $\xoverline{\mathbf{F}}$ and on the right by $\xoverline{\mathbf{F}}^\text{T}$, and upon substituting equation \eqref{WCD1} into equation \eqref{WGCDDev} and equation \eqref{WGCDDev} into equation \eqref{Dev}${}_2$, we obtain
\begin{equation}
p_0\xoverline{\mathbf{B}}=\mu\xoverline{\mathbf{F}}\mathbf{B}_0\xoverline{\mathbf{F}}^\text{T}-\xoverline{\mathbf{F}}\boldsymbol{\tau}\xoverline{\mathbf{F}}^\text{T}
\label{phat0}
\end{equation}
and
\begin{equation}
p_1\mathbf{I}=\mu\xoverline{\mathbf{F}}\mathbf{B}_0\xoverline{\mathbf{F}}^\text{T}-p_0\xoverline{\mathbf{B}}+\xoverline{p}\mathbf{I}-\xoverline{\mathbf{F}}\boldsymbol{\tau}\xoverline{\mathbf{F}}^\text{T},
\label{pbar0}
\end{equation}
respectively. Then substituting \eqref{phat0} into \eqref{pbar0}, we obtain
\begin{equation}
p_1\mathbf{I}=\xoverline{p}\mathbf{I}~~~~~\Rightarrow~~~~~p_1=\xoverline{p},
\end{equation}
as required.
\section{Tensor Identities}
\label{app:Tensors}

The Cayley-Hamilton theorem allows us to determine which tensors are independent. It states that any $3 \times 3$ tensor $\mathbf A$ satisfies
\be
\mathbf A^3  - I_{\mathbf A_1} \mathbf A^2  + I_{\mathbf A_2} \mathbf A - I_{\mathbf A_3} \mathbf I = \mathbf 0,
 \label{eqn:Cayley}
\en
where $ I_{\mathbf A_1}$, $I_{\mathbf A_2}$ and $I_{\mathbf A_3}$ are the invariants of $\mathbf{A}$ analagous to $I_{\vec\tau_1}$, $I_{\vec\tau_2}$ and $I_{\vec\tau_3}$ for $\vec\tau$. From equation~\eqref{eqn:Cayley}, we can see that any power of $\vec \tau$ higher than two can be rewritten in terms of $\vec \tau^2$, $\vec \tau$, $\mathbf I$ and the invariants $I_{\vec\tau_1}$, $I_{\vec\tau_2}$ and $I_{\vec\tau_3}$.

 We will now show that $\tr(\vec \tau^2 \vec \varepsilon)$ and $\vec \tau \vec \varepsilon \vec \tau$,  $\vec \tau^2 \vec \varepsilon \vec \tau + \vec \tau \vec \varepsilon \vec \tau^2$  can be written as combinations of terms already present in equation~\eqref{eqn:CauchyStressLinear}. First substitute $\mathbf{A}=\vec \varepsilon + \gamma \vec \tau$ in equation~\eqref{eqn:Cayley}, where $\gamma$ is an arbitrary scalar. Since the resulting equation must hold for every $\gamma$, each coefficient multiplying a different power of $\gamma$ must be zero individually. The term multiplying $\gamma^2$ is given by
\begin{multline}
\vec \tau \vec \varepsilon \vec \tau +\vec \varepsilon \vec \tau^2 + \vec \tau^2 \vec \varepsilon -(\vec \varepsilon \vec \tau +\vec \tau \vec \varepsilon)I_{\vec\tau_1} - \vec \tau^2 \tr \vec \varepsilon +\vec \tau (I_{\vec\tau_1} \tr \vec \varepsilon - \tr(\vec \varepsilon \vec \tau) ) +\vec \varepsilon I_{\vec\tau_2}
\\
+ \mathbf I (I_{\vec\tau_1}\tr(\vec \tau \vec \varepsilon) -I_{\vec\tau_2} \tr \vec \varepsilon - \tr (\vec \varepsilon \vec \tau^2)  ) =0.
\label{eqn:CayleyCrossed}
\end{multline}
By taking the trace of both sides of this equation (and using the properties $\tr(\mathbf A + \mathbf B)= \tr \mathbf A + \tr \mathbf B$ and $\tr(\mathbf A \mathbf B) = \tr( \mathbf B \mathbf A)$) we establish that $\tr(\vec \tau^2 \vec \varepsilon)$ is indeed a combination of the terms already present in equation~\eqref{eqn:CauchyStressLinear}. The same can then be said for $\vec \tau \vec \varepsilon \vec \tau$ directly from equation~\eqref{eqn:CayleyCrossed},
and for $\vec \tau^2 \vec \varepsilon \vec \tau + \vec \tau \vec \varepsilon \vec \tau^2$ by multiplying equation~\eqref{eqn:CayleyCrossed} on the left by $\vec \tau$.

\section{The entries of the matrix $\mathbf{M}$}
\label{App:M}
The entries of the matrix $\mathbf{M}$ are as follows:
\begin{equation}
 \text{M}_{11}=\alpha_1+3\alpha_2+\alpha_3\tr\vec\tau,\qquad \text{M}_{12}=\alpha_2\tr\vec\tau+\alpha_3\tr(\vec\tau^2)+2\alpha_9I_{\vec\tau_3},
\end{equation}
\begin{equation}
 \text{M}_{13}=\alpha_2\tr(\vec\tau^2)+\alpha_3\tr(\vec\tau^3)+2\alpha_8I_{\vec\tau_3}+2\alpha_9I_{\vec\tau_1}I_{\vec\tau_3},
\end{equation}
\begin{equation}
 \text{M}_{21}=3(\alpha_3+1)+\alpha_6\tr\vec\tau+2\alpha_8,
\end{equation}
\begin{equation}
 \text{M}_{22}=\alpha_1+(\alpha_3+1)\tr\vec\tau+\alpha_6\tr(\vec\tau^2)-2\alpha_9I_{\vec\tau_2},
\end{equation}
\begin{equation}
 \text{M}_{23}=(\alpha_3+1)\tr(\vec\tau^2)+\alpha_6\tr(\vec\tau^3)-2\alpha_8I_{\vec\tau_2}+2\alpha_9(I_{\vec\tau_3}-I_{\vec\tau_1}I_{\vec\tau_2}),
\end{equation}
\begin{equation}
 \text{M}_{31}=2\alpha_9,\qquad \text{M}_{32}=2\alpha_8+2\alpha_9\tr\vec\tau,
\end{equation}
\begin{equation}
 \text{M}_{33}=\alpha_1+2\alpha_8\tr\vec\tau+2\alpha_9\tr(\vec\tau^2).
\end{equation}

\end{document}